\documentclass[preprint,astrosymb]{aastex631}

\usepackage{hyperref}
\usepackage{float}
\usepackage{amsmath}
\usepackage{mathtools}
\usepackage{booktabs}
\usepackage{graphicx}

\newcommand\ujy{\,$\mu$Jy}
\newcommand{\Pspi}{P_\textrm{SPI}}
\newcommand{\abar}{\bar{\alpha}}

\newcommand{\Prot}{P_\textrm{rot}}
\newcommand{\Porb}{P_\textrm{orb}}
\newcommand{\vorb}{v_\textrm{orb}}
\newcommand{\vwind}{v_\textrm{wind}}
\newcommand{\fgeo}{f_\textrm{geo}}
\newcommand{\Rmag}{R_\textrm{mag}}
\newcommand{\Reff}{R_\textrm{eff}}
\newcommand{\Raw}{R_\textrm{AW}}
\newcommand{\Msun}{M_\odot}
\newcommand{\Rsun}{R_\odot}
\newcommand{\Lsun}{L_\odot}
\newcommand{\Bzb}{B_\textrm{ZB}}
\newcommand{\Bzdi}{B_\textrm{ZDI}}

\accepted{8 May 2026}
\submitjournal{ApJ}

\shorttitle{SPI Radio Search}
\shortauthors{Villadsen et al.}
\graphicspath{{./}{figures/}}

\begin{document}

\title{Upper Limits on Planet-Induced GHz Radio Emission from Inactive M~Dwarfs}




\correspondingauthor{Jackie Villadsen}
\email{j.villadsen@bucknell.edu}

\author[0000-0003-3924-243X]{Jackie Villadsen}
\affil{Bucknell University \\
One Dent Drive, Lewisburg, PA 17837, USA}
\affiliation{St.~Mary's College of Maryland \\
47645 College Dr., St. Mary's City, MD 20686, USA}
\affil{Vassar College \\
124 Raymond Avenue, Poughkeepsie, New York 12604, USA}

\author{Carter Russell}
\affiliation{St.~Mary's College of Maryland \\
47645 College Dr., St. Mary's City, MD 20686, USA}


\author{Luna Guerrero}
\affil{Vassar College \\
124 Raymond Avenue, Poughkeepsie, New York 12604, USA}

\author[0009-0005-9484-7732]{Ethan Harvie}
\affil{Bucknell University \\
One Dent Drive, Lewisburg, PA 17837, USA}

\author{Ariana Watson}
\affil{Bucknell University \\
One Dent Drive, Lewisburg, PA 17837, USA}


\author{Arjun Anand}
\affil{Bucknell University \\
One Dent Drive, Lewisburg, PA 17837, USA}

\author[0000-0002-4489-0135]{John Sebastian Pineda}
\affiliation{Laboratory for Atmospheric and Space Physics\\
1234 Innovation Dr, Boulder, CO 80303}

\author{Vanessa Moss}
\affiliation{University of Sydney\\
Camperdown NSW 2006, Australia}
\affiliation{ATNF, CSIRO, Space and Astronomy, PO Box 76, Epping, NSW 1710, Australia}

\author{Daniele d'Antonio}
\affiliation{University of Technology, Sydney}
\affiliation{ATNF, CSIRO, Space and Astronomy, PO Box 76, Epping, NSW 1710, Australia}

\author[0009-0002-5450-6683]{Louisa Canepa}
\affiliation{University of New South Wales}
\affiliation{ATNF, CSIRO, Space and Astronomy, PO Box 76, Epping, NSW 1710, Australia}

\author[0009-0004-2905-6515]{E. Cappellazzo}
\affiliation{School of Mathematical and Physical Sciences, Macquarie University, Sydney, NSW 2109, Australia}
\affiliation{Astrophysics and Space Technologies Research Centre, Macquarie University, Sydney, NSW 2109, Australia}
\affiliation{ATNF, CSIRO, Space and Astronomy, PO Box 76, Epping, NSW 1710, Australia}

\author{Andrew Zic}
\affiliation{Macquarie University}
\affiliation{ATNF, CSIRO, Space and Astronomy, PO Box 76, Epping, NSW 1710, Australia}


\begin{abstract}

Nearby short-period exoplanet systems may produce detectable stellar radio emission due to sub-Alfv\'enic star-planet interaction (SPI), but there are no confirmed cases yet. We targeted five slowly-rotating M dwarfs with transiting terrestrial planets, observing at GHz frequencies throughout their sub-day orbital periods.  We did not detect any bursty SPI-like emission, but detected two stars in quiescence: LHS~3844 (unpolarized) and LHS~1678 (circularly polarized).  These detections imply persistent magnetic activity at Gyr ages, especially notable for LHS~1678 given its low photometric variability, and can serve as targets for radio transit experiments. Our SPI non-detections may be due to radio beaming geometry, a sub-GHz maximum emission frequency, or undetectable flux density. If the last case applies, then flux density upper limits constrain the exoplanet magnetosphere. GJ~367~b has the tightest constraints -- no extended magnetosphere and an exoplanet field $<$0.8~G -- although these results depend strongly on unknown stellar wind parameters inferred from stellar rotation period. Due to their small orbital distance, our non-detection systems a priori appear to have more favorable conditions for SPI than most radio-detected SPI candidate systems in the literature, a tension that can either be resolved by favorable wind/geometry conditions on the detected candidates or by a non-SPI (stellar activity) explanation for those candidate detections. Our results favor the approach of sub-GHz searches for radio SPI, especially with the sensitivity of new/upcoming facilities such as MeerKAT, and underscore the need for observational and theoretical work to constrain the magnetized stellar wind parameters.

\end{abstract}

\section{Introduction}

Exoplanet magnetic fields provide a unique tracer of planets' internal structure and modulate their atmospheric loss \citep{brain2024exoB_chapter}.  This valuable tracer is difficult to detect directly, with searches for cyclotron emission from planetary aurorae remaining inconclusive \citep{turner2024tauBoo_followup} and terrestrial fields of $\sim$1~G requiring space-based MHz observatories \citep{burns2021farside,knapp2024go-low}.  In the meantime, indirect detection of exoplanet magnetospheres through star-planet interaction (SPI) may provide initial estimates of exoplanet magnetic field strengths, especially as TESS and ground-based radial velocity experiments rapidly expand the known population of nearby, close-in planets.

Close-in planets move through the stellar wind slower than the local Alfv\'en speed, allowing energy transfer back to the host star along the star-planet magnetic flux tube.  In such sub-Alfv\'enic SPI, the power transferred back to the star depends on the size of the exoplanet magnetosphere and thus the planet's magnetic field strength. The transferred energy should produce orbitally-modulated stellar activity signals. Such orbital modulation has been observed in chromospheric activity lines \citep{cauley2019NatAs} and optical flaring \citep{ilin2025optical_SPI}, and proposed for gyrosynchrotron radio flaring \citep{osten2025v830tau_gyrosynch_SPI} and auroral radio emission.

Sub-Alfv\'enic radio SPI is expected to manifest as bursts of circularly-polarized radio emission analogous to Jupiter's auroral radio bursts caused by Io and Ganymede \citep{zarka2018ganymede}, where the emission mechanism is the electron cyclotron maser (ECM).  The range of cyclotron frequencies along the star-planet flux tube determines the emission band, which should drop off at a maximum ECM cutoff frequency corresponding to the stellar surface field.  ECM amplification causes a beaming pattern shaped like a wide, hollow cone, such that bursts of emission occur at orbital phases when the beam points at Earth \citep{hess2011spi_dynspec,kavanagh2023geometry}.

Solar system-based models \citep{zarka2007spi_model,saur2013} predict that sub-Alfv\'enic SPI from nearby exoplanetary systems may be detectable by current radio facilities at MHz to GHz frequencies \citep{turnpenney2018}.  The search for radio SPI can be divided into two steps: identifying candidate radio-emitting systems, and confirming orbital modulation of the radio signal, which has not been definitively accomplished for any radio SPI candidate to date.  

Testing for orbital periodicity can require hundreds of hours of radio monitoring.  In identifying targets for long-term monitoring, studies may choose to focus on either young magnetically-active stars or older inactive stars.  Young strongly-magnetized stars offer the benefit of a high predicted SPI luminosity, enabling sensitivity to weak planetary fields, which makes the recent radio non-detection of two rapidly-rotating M dwarf planet hosts particularly surprising \citep{ortiz-ceballos2025active_dM_spiSearch}. 
However, even in the case of a radio detection, strong stellar activity also causes a foreground of non-SPI-driven stellar radio bursts.
So far, long programs targeting a young planet-hosting M~dwarf \citep{bloot2024aumic_radio} and G~dwarf \citep{ilin2025radioSPI_youngG} have found no evidence of orbitally-phased radio emission, instead finding that the radio emission is stochastic or phased with the star's rotation, as expected for non-planet-driven stellar activity.

To mitigate the challenge of non-SPI stellar activity, another approach focuses on identifying candidate SPI systems as radio emitters with relatively weak stellar activity.  One method to find low-activity SPI candidates is to use radio surveys to identify stars with anomalously strong radio emission compared to other activity indicators, since the radio emission could indicate an undiscovered SPI-inducing planet \citep[e.g.,][]{vedantham2020, Pope2021ApJ...919L..10P, frail2025quiet_dK_radio}.  This blind-survey method has exciting potential as a novel exoplanet detection method, although it faces the challenge that initial radio detections may be exceptional events that do not recur in follow-up \citep{narang2024gj1151_followup}. Another method to find low-activity SPI candidates is targeted radio observations of known close-in exoplanet systems with older stars, yielding polarized radio detections on M~dwarfs Prox~Cen \citep{perez-Torres2021}, YZ~Cet \citep{pineda2023yzcet}, GJ~3323 \citep{ortiz2024gj3323_exoradio}, and K+M dwarf HD~189733 \citep{zhang2025hd189733_lowfreq_burst}, and upper limits for M~dwarf planet hosts GJ~486 \citep{pena-monino2025gj486_spi} and K2-18 \citep{wandia2026radioSPIsearch_k2_18b}. In a promising conjunction of the blind-survey and targeted-exosystem approaches, a recent broad stellar survey with LOFAR found a radio burst with Jupiter-Io-like morphology, which originated from an M dwarf with a known planet \citep{tasse2026lofar_beamforming_SPI_search}.

In this work, we expand the search for low-activity, sub-Alfv\'enic radio SPI candidates, through long-duration, GHz observations of 5 recently-discovered close-in exoplanet systems around slowly-rotating M~dwarfs.  In Section~\ref{sec:targets}, we present the 5 targets and a comparison sample from the literature.  In Sections~\ref{sec:obs}-\ref{sec:results}, we describe the observations, data analysis, and results, including quiescent detections of 2 systems but no detections of SPI-like polarized bursts.  In Section~\ref{sec:wind}, we prepare to interpret the non-detections by setting up a wind model for our targets based on rotation-magnetism and rotation-mass loss correlations. In Section~\ref{sec:nondetections}, we consider 4 possible causes of the SPI non-detections, one of which (low luminosity) we expand on in Section~\ref{sec:spi_model} by modeling our radio upper limits as constraints on exoplanet magnetic field, with the conclusions summarized in Section~\ref{sec:conclusions}. 
      
\section{Targets} \label{sec:targets}

\begin{deluxetable*}{l|llllll|llLl|lll}
\tablewidth{0pt} 
\tablecaption{Target Systems\label{tab:targets}}
\tablehead{
\colhead{} & \multicolumn{6}{c}{\textbf{Stellar Parameters}} & \multicolumn{4}{c}{\textbf{Exoplanet b}} &            \multicolumn{3}{c}{\textbf{Radio Observations}}         \\
\colhead{}  & \colhead{SpT} & \colhead{$M_*$} & \colhead{$R_*$} & \colhead{$d$} & \colhead{$\Prot$} & \colhead{$L_\textrm{bol}$} & \colhead{$R_p$} & \colhead{$\Porb$} & \colhead{$a$} & \colhead{Ref} & \colhead{$F_\nu$}          & \colhead{Freq.}  & \colhead{Ref}     \\
\colhead{}   & \colhead{} & \colhead{($\Msun$)}               & \colhead{($\Rsun$)}              & \colhead{(pc)}     & \colhead{(d)}    & \colhead{($\Lsun$)}     & \colhead{($R_E$)}            & \colhead{(d)}    & \colhead{($R_*$)} & \colhead{}             & \colhead{(\ujy)}                      & \colhead{(GHz)}   &    }    
\startdata
LTT 3780  & M3.5 & 0.379              & 0.382             & 22.03  & 137  & 1.67$\times10^{-2}$ & 1.35              & 0.768 & 6.69       &  1    & $<$75            & 2-4       &    *    \\
LHS 3844  & M4.5 & 0.151              & 0.189             & 14.878 & 128  & 2.72$\times10^{-3}$ & 1.303             & 0.463 & 7.09       &  2    & $<$350           & 1.1-3.1   &   *   \\
GJ 367    & M1 & 0.454              & 0.457             & 9.419  & 48   & 2.88$\times10^{-2}$ & 0.718             & 0.322 & 3.32         &   3   & $<$330           & 1.1-3.1   &   *   \\
LHS 1678  & M2 & 0.345              & 0.329             & 19.865 & 64   & 1.45$\times10^{-2}$ & 0.7               & 0.860 & 8.11         &   4   & $<$610           & 1.1-3.1   &   *   \\
GJ 1252   & M3 & 0.381              & 0.391             & 20.385 & 64   & 1.96$\times10^{-2}$ & 1.193             & 0.518 & 5.03         &   5   & $<$460           & 1.1-3.1   &   *   \\
\hline
YZ Cet    & M4.5 & 0.137              & 0.163             & 3.717  & 63.5 & 2.25$\times10^{-3}$ & 0.89              & 2.02 & 21.3           &  6,A,$\dagger$   & 470-1070                 & 2-4       &   A     \\
GJ 486    & M3.5 & 0.333              & 0.339             & 8.079  & 49.9 & 1.21$\times10^{-2}$ & 1.305             & 1.47 & 11.1           &  7   & $<$216           & 0.55-0.75 &  B   \\
Prox Cen  & M5.5 & 0.12               & 0.141             & 1.301  & 89   & 1.55$\times10^{-3}$ & 1.1               & 11.2 & 73.7           &  8,$\dagger$   & 200-5000                  & 1.1-3.1   &   C   \\
GJ 3323   & M4.5 & 0.164              & 0.119             & 5.375  & 99   & 2.70$\times10^{-3}$ & 1.26              & 5.36 & 59.3           &  9,$\dagger$   & 86$\pm$10 & 4-8       &  D      \\
\hline
Reference & M2.5 & 0.3                & 0.3               & 10     & 100  & 1.00$\times10^{-2}$ & 1                 & 1   & 9.4            &  10   & $<$100           & 0-2       &       
\enddata
\tablecomments{Five target systems observed in this paper, then 4 comparison systems from the literature, and one hypothetical reference system. \textit{Star/exoplanet references:} Distances from \cite{Gaia2020yCat.1350....0G}. 
(1) \cite{nowak2020ltt3780}
with $\Prot$ from \cite{sairam2025ltt3780_xray}
(2) \cite{vanderspek2019} (3) \cite{lam2021gj367} (4) \cite{silverstein2022lhs1678} (5) \cite{shporer2020gj1252} (6) \cite{stock2020yzcet} (7) \cite{caballero2022gj486} (8) \cite{anglada2016proxcen_b} (9) \cite{astudillo2017gj3323} (10) Typical 0.3$\Msun$ spectral type and luminosity from \cite{baraffe1996spT_mass}.  ($^\dagger$) We assume Earth density and minimum mass for non-transiting systems.
\textit{Radio references:} (A) \cite{pineda2023yzcet}, see also 0.55-0.9~GHz observations by \cite{trigilio2023yzcet_arxiv} (B) \cite{pena-monino2025gj486_spi}, with their $3\sigma$ Stokes~V sensitivity to 5 minutes to match our observations (C) \cite{perez-Torres2021} (D) \cite{ortiz2024gj3323_exoradio} (*) This work, $3\sigma$ Stokes~V 5-minute upper limits.}
\end{deluxetable*}

We have targeted five M dwarfs known to host ultra-short-period terrestrial exoplanets (Table~\ref{tab:targets}).  None of these targets have prior targeted radio observations in the literature.  For multi-planet systems, the table lists only the innermost planet, since the strong dependence of SPI power on orbital distance ($\Pspi \propto a^{-4}$ to $a^{-5}$; Section~\ref{sec:Pspi_unmag}) favors detecting close-in planets.  We also compare to 4 slowly-rotating M~dwarf terrestrial planet hosts with literature radio observations, and a hypothetical reference system with easily scalable basic parameters.  Our 5 target systems were all discovered via the transit method and thus have radii available, whereas for RV-discovered comparison systems YZ~Cet~b, Prox~Cen~b, and GJ~3323~b we assume Earth density and edge-on orientation: $R_p = (M_p \sin i)^{1/3}$ in Earth units.

The nearness to Earth and short orbits of our targets make them interesting follow-up targets even though they are too hot to be habitable, for instance LHS~3844~b has potential for JWST  characterization \citep{Whittaker2022AJ....164..258W}. 
However, at a reduced semi-major axis of $a/R_*\sim$ 3-8, our targets should be tidally locked in circularized orbits, with stellar irradiation sufficient to have evaporated any Earth-like atmospheres \citep[e.g., GJ~367;][]{poppenhaeger2024gj367_xray}. Tidal locking, circularization, and lack of atmosphere are supported by IR secondary eclipse observations of GJ~1252 \citep{crossfield2022gj1252b_noAtmosphere} and LTT~3780 \citep{allen2025ltt3780_jwstHotRock}, and IR orbital phase curves for GJ~367~b \citep{zhang2024gj367} and LHS~3844~b \citep{kreidberg2019,lyu2024lhs3844b_tidallyLocked}, with the latter's atmosphere also undetected in ground-based transit spectroscopy \citep{DiamondLowe2020AJ....160..188D}.  The planets likely have volatile-poor, solid surfaces \citep[e.g.,][]{Kane2020PSJ.....1...36K}, but alternative proposed interpretations revive the possibility of an atmosphere, including explaining phase curves with a thick atmosphere with night-side clouds \citep{powell2024phasecurve_clouds} and a novel form of tectonics driven by the dayside/nightside temperature contrast \citep{Meier2021ApJ...908L..48M}.  Overall, the preponderance of evidence suggests that our target planets are unlikely to have an ionosphere that can form an induced magnetosphere similar to Io's.  However, if they have a strong extended magnetosphere above the planet's surface similar to Ganymede, then strong SPI interaction can occur without an ionosphere \citep{saur2013}.

All of our target stars are slowly rotating early-to-mid M dwarfs with weak stellar magnetic activity and low flare rates. Slow rotators minimize the risk of contamination by non-SPI flaring \citep{Pope2021ApJ...919L..10P} and rotationally-powered aurorae \citep{vedantham2020}. The rotation periods in Table~\ref{tab:targets} are 
photometric.  Studies have found weak activity indicators for all of our targets consistent with slow rotation: $\log R'_{HK}=-5.6$, non-detection in H$\alpha$, and no flares in one TESS sector for LTT~3780 \citep{Cloutier2020AJ....160....3C,Jeffers2018A&A...614A..76J,Pope2021ApJ...919L..10P}; undetected or unsaturated H$\alpha$, CaII, and HeI D for LHS~3844 with only one TESS flare \citep{Medina2020ApJ...905..107M}; $\log R'_{HK}=-5.2$ for GJ~367 \citep{lam2021gj367}; absorption in H$\alpha$ for LHS~1678 \citep{silverstein2022lhs1678}; and $\log R'_{HK}=$-5.4 and H$\alpha$ absorption for GJ~1252 \citep{astudillo2017gj3323,shporer2020gj1252}.  Perhaps the most notably quiet target is LHS~1678, which showed no spot modulation or flares in 4 months of TESS \citep{silverstein2024_lhs1678d}, and which \cite{kar2024AJphot_var} identify as the most photometrically stable of 32 M dwarfs monitored for multiple years.

Compared to younger M dwarfs, the predicted SPI power from slow rotators is reduced due to their weakened stellar magnetic fields and winds.  However, compared to FGK stars, even slowly-rotating M dwarfs can have relatively strong global magnetic fields, increasing both SPI power and the emitted cyclotron frequency (see Section~\ref{sec:nondetections}).  For example, the Sun and Prox Cen, respectively, have large-scale magnetic fields of $\sim$10~G \citep{Vidotto2016MNRAS.459.1533V} and 200~G \citep{Klein2021MNRAS.500.1844K}.  

\begin{deluxetable*}{llllll}
\tablewidth{0pt} 
\tablecaption{Observations\label{tab:obs}}
\tablehead{
\colhead{}   & \colhead{Telescope \& Band} & \colhead{Project}
& \colhead{Epochs}  & \colhead{$\Porb$}
& \colhead{Orbital} \\
& & & \colhead{(Duration in hr)} & \colhead{(hr)} & \colhead{Coverage}
}
\startdata
LTT 3780 & VLA 2-4 GHz       & 19B-274 & 2020-01-09 (6.3), 01-10 (6.3), 01-11 (6.3) & 18.4 &  96.5\%                     \\
LHS 3844 & ATCA 1.1-3.1 GHz  & C3303   & 2019-06-21 (11.8), 06-26 (11.4)              & 11.1 &   100\%                    \\
GJ 367   & ATCA 1.1-3.1 GHz  & C1726,  & 2022-01-19 (9.5), 09-13 (7.3),             &  7.7 &    100\%                   \\
         &                   & C3303   &  09-14 (8.2), 09-15 (8.0), 09-24 (7.6)         &      &                       \\
LHS 1678 & ATCA 1.1-3.1 GHz  & C3303   & 2022-09-23 (7.6), 09-26 (9.5), 09-27 (10.2)    & 20.6 &  81.4\%                     \\
GJ 1252  & ATCA 1.1-3.1 GHz  & C3303   & 2022-09-28 (12.9), 09-29 (12.7)              & 12.4 &   98.8\%                   
\enddata
\tablecomments{The final column lists the percentage of orbital phase that was observed at least once.}
\end{deluxetable*}

\section{Observations}
\label{sec:obs}

Table~\ref{tab:obs} summarizes the observations.  Four of the targets were observed with the Australia Telescope Compact Array (ATCA) at 1.1-3.1 GHz, in projects C1726 (PI: Hollow) and C3303 (PI: Pineda). LTT~3780 was observed with the Karl G.~Jansky Very Large Array (VLA) at 2-4 GHz, in project 19B-274 (PI: Villadsen).  The arrays had 5-6 antennas per observation date (ATCA) or 26-27 (VLA). The observation durations in the table count from the beginning of the first to the end of the last target scan.  These durations slightly exceed the true time on source due to calibrator observations, which occurred for 1-3 minutes every 20-60 minutes.  Each system was observed across 80-100\% of orbital phases at least once, with multi-date observations sampling across different orbits in order to achieve near-complete phase coverage.  

LTT~3780 is the primary star in a $16''$ visual binary with M5 dwarf LP~729-55 \citep{nowak2020ltt3780}.  We did not find records of companions for any other targets.  Since we are searching for a strongly circularly-polarized signal, the probability of source confusion is low.

\section{Data Analysis}
\label{sec:data_analysis}

\noindent For LHS~3844, we flagged and calibrated the ATCA data in Miriad. For the other ATCA targets, we flagged and calibrated in CASA. We performed polarization calibration for the ATCA data in order to use the cross-hand X and Y signals to measure circular polarization. For LTT~3780, we flagged and calibrated the VLA data using the VLA pipeline in the CASA software package \citep{casa2007}. The observations typically achieved an effective bandwidth of $\sim$1.4 GHz due to radio frequency interference (RFI).

For all targets, we used CASA to generate both total intensity (Stokes~I) and circular polarization (Stokes~V) images using the entire observation duration. Nearby dates were combined, but for GJ 367, the January and September epochs were imaged separately, as were the Jun 21 and 26 dates for LHS~3844.  We deconvolved Stokes~I images in CASA using the \texttt{tclean} task with W-projection, multifrequency synthesis with 2-4 Taylor terms (by default 2, up to 4 if a bright background source was near the primary beam null), and multi-scale model components. This imaging serves a dual purpose: searching for quiescent (time-averaged) stellar emission in the image plane, and creating model visibilities for background sources in the field of view.  The ATCA Stokes~I images did not initially reach near target sensitivity, so we performed 1-2 rounds of phase-only self-calibration. 

Before making time series, we shifted the phase center of the visibility data to the stars' known location adjusted for proper motion \citep{Gaia2020yCat.1350....0G} using the SIMBAD database, unless it was already aligned to within a small fraction of a synthesized beam. The phase centers used are: LTT~3780 at $(\alpha,\delta)=(10^h 18^m 34\fs 672, -11\arcdeg 43\arcmin 05\farcs 206)$, LHS~3844 at $(22^h 41^m 59\fs 340, -69\arcdeg 10\arcmin 22\farcs 496)$, GJ~367 at $(09^h 44^m 28\fs 863, -45\arcdeg 46\arcmin 48\farcs 244)$ in January 2022 and $(09^h 44^m 28\fs 833, -45\arcdeg 46\arcmin 48\farcs 650)$ in September 2022, LHS~1678 at $(04^h 32^m 43\fs 106, -39\arcdeg 47\arcmin 34\farcs 110)$, and GJ~1252 at $(20^h 27^m 43\fs 243, -56\arcdeg 27\arcmin 53\farcs 098)$.  None of the targets are known binary systems.  We did not adjust for parallax motion (0.05''-0.1''), since it is much smaller than the synthesized beam size of $\gtrsim5\arcsec$.

We made time series plots for the locations of the target stars in both Stokes I and V. To do so, we first modeled and removed background sources from the data. The imaging process (\texttt{tclean}) created model visibilities for the background sources, using masking to omit the location of the star from the model. Then, we used \texttt{uvsub} to subtract that model from the visibility data, leaving residual visibilities containing only the target star, plus thermal noise, RFI, and traces of imperfectly subtracted background sources. We then averaged the residual visibilities across all baselines using \texttt{plotms}, creating a complex time series for the phase center.  The real component is the time series for the star, and the imaginary component should not contain stellar emission and thus provides an estimate of noise levels.

\begin{figure}[ht!]
\plottwo{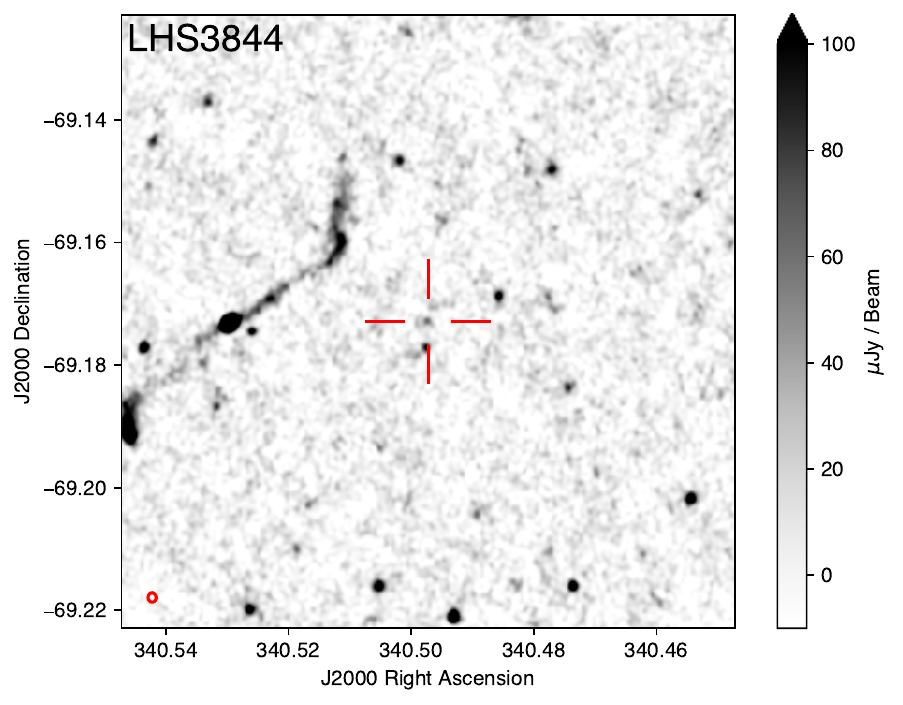}{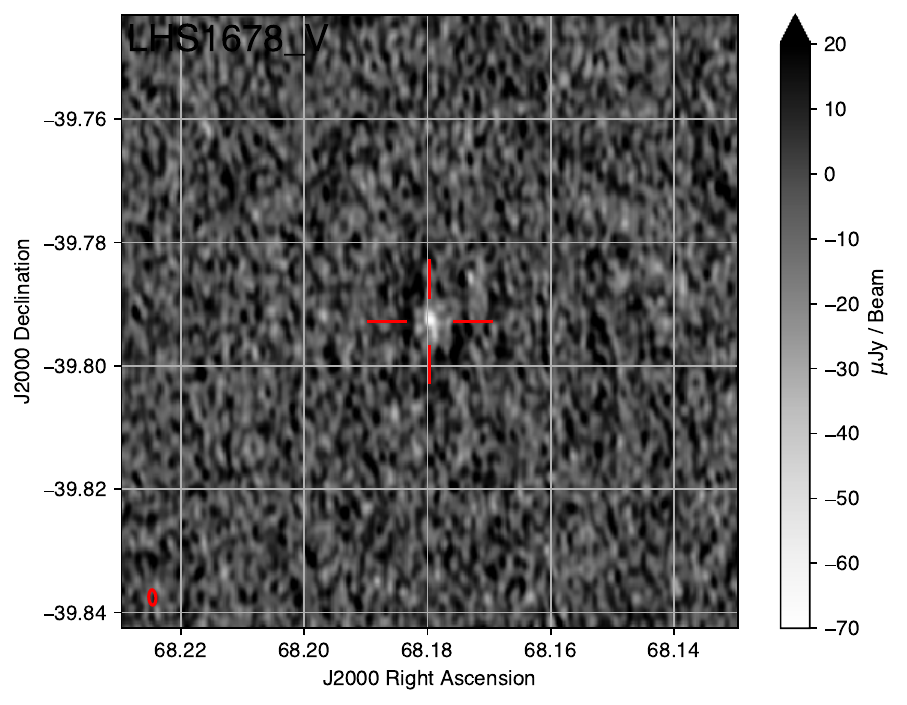}
\caption{\textit{(left)} ATCA 1.1-3.1~GHz Stokes~I image of LHS~3844 during its first epoch.  The star is detected in Stokes~I at $63\pm12$~\ujy, where the uncertainty is estimated from the image RMS in empty regions near the star's location.  The star is undetected in Stokes~V. \textit{(right)} Stokes~V image of LHS~1678 during its three epochs combined.  The star is detected in Stokes~V at -70$\pm$9~uJy, and undetected in Stokes~I (Appendix~\ref{app:images}).
\label{fig:detection_images}}
\end{figure}

\section{Results}
\label{sec:results}

Appendices~\ref{app:images} and \ref{app:tseries} present the images and time series, respectively, for all targets. None of the stars were detected in the time series, using a detection criterion of 2 adjacent points in the time series exceeding three times the standard deviation $\sigma$.  We binned the time series to integration times of 1, 3, 5, 10, and 15-30 minutes (maximum depending on scan length), with no multi-point detections at any time binning. While the expected burst duration is quite uncertain due to its strong dependence on orbital and magnetic geometry \citep{kavanagh2023geometry}, an emission cone thickness of 1\degr~(consistent with Jupiter-Io) should sweep across the line of sight in a time of order minutes for the targets' short orbital periods.

Throughout the remainder of the paper, we use $3\sigma$ upper limits on the flux density based on the standard deviation of the Stokes~V time series with 5-minute time bins (shown in Appendix~\ref{app:tseries}).  The 3$\sigma$ upper limits are summarized in Table~\ref{tab:targets}; for multi-epoch observations, the table presents the standard deviation of all epochs combined, except for GJ~367 where each epoch covers almost the full orbital period so we report the most sensitive epoch's limits.  The CABB sensitivity calculator predicts $\sigma=$60 $\mu$Jy sensitivity in 5 minutes an ATCA 1.1-3.1~GHz with typical weather, 6 antennas, and RFI removed. Our measured Stokes~V sensitivities range from 90-200~\ujy, due to a missing antenna in some epochs plus poor weather, heavy RFI, and/or sidelobes of bright off-axis background sources leaking into Stokes~V.

Two targets were detected in the image plane in images of the full duration of one or more epochs. LHS~3844 was detected in Stokes~I with a flux density of $63\pm12$\ujy~in the first of two epochs (Figure~\ref{fig:detection_images}) and $56\pm17$\ujy~in the second epoch (Appendix~\ref{app:images}), although there is more visible sidelobe contamination around the source in the second epoch. LHS~1678 is detected in Stokes~V in an image combining all 3 epochs (Figure~\ref{fig:detection_images}), at a flux density of -70$\pm$9\ujy. While this is a $7\sigma$ detection and there are no known background sources at this location, it has a somewhat non-point-like shape that may be due to time-variable poor phase stability or polarization calibration, so we consider this a tentative detection. Imaging single epochs did not improve the signal-to-noise ratio of the LHS~1678 images.  The Stokes~I flux density at the star's location (no clear source) is $63\pm40$\ujy, estimated from the intensity value at the center pixel in the image and the image RMS near that location. The nearby Stokes~I image RMS is large due to a nearby extended source with unclear boundaries, which is clearly offset from the Stokes~V source but whose edges may overlap this location. 

\section{Stellar Wind Model} \label{sec:wind}

The possible causes of non-detections (Section~\ref{sec:nondetections}) depend on the magnetic field and density of the stellar wind, but our targets do not have observational constraints on their wind properties.  In this section, we first predict the stellar wind base conditions: stellar magnetic field, mass loss rate, and temperature. We then use a radial wind model to predict circumplanetary wind density and field strength, shown in Figure~\ref{fig:wind_model}.  We conclude this section with a rough estimate of the uncertainties on our predictions.

\subsection{Wind base conditions: Large-scale photospheric magnetic field}

Our target planets orbit at $3-8 R_*$, where the effects of small-scale surface fields are negligible and large-scale field dominates the wind properties \citep{lang2014smallscale}, so we seek to predict the large-scale surface field that would be detectable by Zeeman Doppler Imaging (ZDI), $\Bzdi$.  While rotation-$\Bzdi$ correlations exist for more massive stars \citep{vidotto2014rot_vs_Bzdi}, slowly-rotating M dwarfs exceed these correlations by an order of magnitude \citep{lehmann2024slowrot_dM_zdi,see2025slowrot_dM_Bfields}, but new rotation-$\Bzdi$ correlations are not yet available for this mass/rotation parameter space as ZDI characterization of these targets is new and ongoing \citep{donati2023dM_spectropol_monitoring}. Instead, to estimate $\Bzdi$, we first use rotation to predict the average surface field detectable by Zeeman Broadening (ZB), which includes small-scale features.

We apply the observational ZB relationship of \cite{reiners2022dM_ZB_correlation} [R22] (their Eq.~2). All of our targets are slow rotators with Ro $>$ 0.13 (Table~\ref{tab:wind}), so the relationship predicts:
\begin{equation}
\Bzb = (8570~\textrm{G}) R_*^{-2} \Prot^{-1.25},
\label{eq:B_ZB}
\end{equation}
where $R_*$ is in solar units and $\Prot$ in days.  This formula predicts fields of 130-600~G for our targets (Table~\ref{tab:wind}) and 600-2000~G for our comparison sample since they are later spectral type.

We then translate from the predicted ZB field, to estimate the magnitude of the large-scale field $\Bzdi$ observable by ZDI:
\begin{equation}
    \Bzdi = f_\textrm{ZDI} ~ \Bzb
    \label{eq:B_ZDI}
\end{equation}
using the typical ratios measured by \cite{reiners2009frac_BV_BI} of $f_\textrm{ZDI} = |B_V/B_I|\sim$ 6\% for partially convective stars and 14\% for fully convective stars.  We use a partially/fully-convective boundary of $M_* = 0.35 \Msun$.
This calculation predicts large-scale surface fields of 8-80~G.  We assume that the ratio $\Bzdi/\Bzb$ is the same as $B_V/B_I$, which may underestimate $\Bzdi$ since $B_V$ is reduced by projection effects. However, the error introduced is small compared to other uncertainties, and \cite{wulff2026bzb_vs_bzdi} find that the dipole field of slowly-rotating M dwarfs \citep{see2025slowrot_dM_Bfields,lehmann2024slowrot_dM_zdi} is typically about 10\% of $\Bzb$, matching our 6-14\% ratio.   Based on the few ZDI observations of slowly-rotating M~dwarfs, the large-scale field is probably almost entirely dipolar \citep{lehmann2024slowrot_dM_zdi} [hereafter L24].

While none of our 5 targets have ZB or ZDI measurements, some of the comparison sample do, but we use the rotation-based magnetic field predictions throughout this work for simplicity.

\begin{deluxetable*}{l|ll|lllll}
\tablewidth{0pt} 
\tablecaption{Predicted stellar wind parameters\label{tab:wind}}
\tablehead{
\colhead{} & \colhead{$\Bzb$}  & \colhead{$\Bzdi$} & \colhead{Ro} & \colhead{$L_X$} & \colhead{$L_X$}
& \colhead{$\dot{M}/A$} & \colhead{$\dot{M}$} \\
\colhead{} & \colhead{(G)} & \colhead{(G)} & & \colhead{(erg\,s$^{-1}$)} & \colhead{ref} & \colhead{(solar)} & \colhead{($\dot{M}_\odot$)}
}
\startdata
LTT 3780 & 125 & 7.5 & 2.14 & 4.3$\times10^{25}$ & * & 0.27  & 0.039 \\
LHS 3844 & 557 & 78 & 0.99 & 5.6$\times10^{25}$ & * & 0.97  & 0.035 \\
GJ 367   & 325 & 19 & 0.93 & 5.0$\times10^{26}$ & 1 & 1.34  & 0.28  \\
LHS 1678 & 437 & 61 & 0.91 & 3.8$\times10^{26}$ & * & 1.81  & 0.20  \\
GJ 1252  & 310 & 19 & 1.01 & 3.9$\times10^{26}$ & * & 1.41  & 0.21  \\
\hline
YZ Cet   & 1800 & 252 & 0.47 & 6.4$\times10^{26}$ & 2 & 7.95  & 0.21  \\
GJ 486   & 562 & 78 & 0.68 & 6.3$\times10^{25}$ & 3 & 0.43  & 0.050 \\
Prox Cen & 1580 & 221 & 0.62 & 1.7$\times10^{27}$ & 4 & 19.4 & 0.42 \\
GJ 3323  & 1940 & 271 & 0.80 & 1.9$\times10^{27}$ & 5 & 30.1 & 0.43 \\
\hline
Reference & 301 & 42 & 1.23 & 1.1$\times10^{26}$ & * & 0.82 & 0.073
\enddata
\tablecomments{References for $L_X$: (*) Estimated from Rossby number ($\Prot$ in Table~\ref{tab:targets}) using \cite{wright2011rot_activity_Lx}. The LHS~3844 and LTT~3780 predictions are consistent with the observational limits of $L_X<2.9\times10^{26}$ and $5.0\times10^{26}$~erg\,s$^{-1}$ respectively \citep{DiamondLowe2021AJ....162...10D,sairam2025ltt3780_xray}. X-ray detections: (1) \cite{poppenhaeger2024gj367_xray} (2) \cite{stelzer2013dM_xray} (3) \cite{diamondlowe2024gj486_xray} (4) \cite{schmitt2004xray_survey}
(5) \cite{freund2024erosita_xray_stars}.}
\end{deluxetable*}

\subsection{Wind base conditions: Mass loss rate}
\label{sec:Mdot}

None of our targets have wind measurements.  To estimate their mass loss rates, we employ the power-law fit of \cite{wood2021dM_winds} for a population of main-sequence stars including about half M dwarfs, which relates astrospheric and slingshot prominence wind measurements to X-ray fluxes.  Their correlation (their Figure~10) between mass loss per unit surface area $\dot{M}/A$ (solar units) and surface X-ray flux $F_\textrm{X,surf}$ (erg\,cm$^{-2}$\,s$^{-1}$) is:
\begin{equation}
\dot{M}/A = 10^C \, F_\textrm{X,surf}^\beta
\end{equation}
where the fit coefficients are $C=-3.41\pm0.26$ and $\beta=0.77\pm0.04$ (B.~Wood, private communication).

Table~\ref{tab:wind} lists our targets' X-ray luminosities $L_X$ and predicted mass loss rates.  When available, we use published X-ray luminosities for our targets and comparison sample.  However, likely due to their weak activity, 4 of our 5 targets have no X-ray detections, so we instead employ a rotation-activity relationship to estimate $L_X$, similar to the approach of \cite{cilley2024xray_predict}. We use the empirical formula from \cite{wright2018rot_activity_fully_conv} for M-dwarf convective turnover time:
\begin{equation}
\log_{10} \tau=2.33 - 1.5M_* + 0.31M_*^2
\end{equation}
where $M_*$ is in solar units.  All of our targets are unsaturated due to their slow rotation, with a Rossby number $Ro > Ro_\textrm{sat}=0.16$. We use the rotation-activity relation of \cite{wright2011rot_activity_Lx} in the unsaturated regime, which \cite{wright2018rot_activity_fully_conv} found also applies to fully convective M dwarfs:
\begin{equation}
    L_X/L_\textrm{bol} = C \, Ro^\beta
\end{equation}
where $\beta=-2.7$ and $(L_X / L_\textrm{bol})_\textrm{sat} = 10^{-3.13}$ give $C=(L_X / L_\textrm{bol})_\textrm{sat} Ro_\textrm{sat}^{-\beta} = 5.26\times10^{-16}$.

We predict mass loss per area similar to solar values for our targets (0.3-2 times solar; Table~\ref{tab:wind}), due to their observed/expected weak coronal activity.

\subsection{Wind base conditions: Temperature} \label{sec:wind_temp}

We model the stellar wind as an isothermal Parker wind.  To do so, we estimate the wind temperature from X-ray coronal temperatures.
GJ~367, our only X-ray-detected target, has a coronal temperature of 1.82~MK \citep{poppenhaeger2024gj367_xray}. \cite{brown2023coronal_temp} [B23] found that the coronal temperature for dozens of M dwarfs (mostly slow rotators) was in the range 2-6~MK, with many stars clustered around 3~MK, suggesting that the typical coronal temperature of our targets may be around 3~MK. However, many of B23's X-ray data could only be fit with a single temperature component, which is dominated by dense, hot active regions, and which may thus overestimate the wind temperature in open-field regions. For example, \cite{ofionnagain2018youngSun_wind} [OV18] find that the relation between rotation and average X-ray coronal temperature from \cite{johnstone2015spindown_Mdot} predicts a coronal temperature for the Sun that is 1.36 times higher than the basal solar wind temperature of 1.5~MK needed for their polytropic model to match the solar wind.

To mitigate the risk of overestimating wind temperatures based on an average coronal temperature, we can also look at the lower temperature component in multi-temperature fits to M dwarf coronal X-ray observations. This approach is consistent with Sun-as-a-star X-ray observations, which have a dominant coronal emission component at 1~MK \citep{peres2000sunAsStar_Xray}, agreeing with a 1.1-MK isothermal fit to the slow solar wind \citep{sheeley1997solarwind}.
For the 5 of 23 stars in B23 that have a 2 or 3-temperature fit to the X-ray while in a quiescent or low state, the low-temperature component ranges from 1.4-1.7 MK.  We choose a wind temperature that lies between (3~MK)/1.36 (based on the OV18 scale factor and typical coronal temperatures from B23) and 1.5~MK (based on a typical low-temperature component in B23), and thus we assume a typical wind temperature of 2~MK for our targets.  Since the average coronal temperatures in B23 vary from 2-6 MK, roughly a factor of 2 around the typical value of 3~MK, we also assume a factor of 2 uncertainty in wind temperature due to the variation between stars. 

\subsection{Wind magnitude at planet: Stellar magnetic field} \label{sec:B_profile}

The wind magnetic field strength at the planet depends on whether the stellar field is open or closed at that distance.  \cite{vidotto2014dM_ZDI_winds} performed ZDI-grounded MHD wind simulations of 6 active early M dwarfs, and found that the ``effective source surface,'' where the field is almost entirely radial and open, occurs at 2.8-4.6$R_*$.  An open stellar field at $\gtrsim 3R_*$ should thus be a reasonable approximation for all targets, even GJ~367~b at 3.3$R_*$.  Since our stars are slow rotators, we also assume the field is purely radial with no azimuthal component, so $B(\tilde{r}) = B_\textrm{open} \tilde{r}^{-2}$, where we define $\tilde{r} = r/R_*$.

How does the large-scale surface field $\Bzdi$ relate to the stellar surface-averaged open radial field $B_\textrm{open}$?  \cite{lehmann2024slowrot_dM_zdi} produced ZDI maps of six slowly-rotating M~dwarfs, typically detecting only the poloidal field, which is dominated by a radial component but also has a weaker meridional component (50\% or less of the radial component).  Since the magnetic field components add in quadrature, the radial component of the ZDI magnetic field matches the overall magnitude to within $\sim$10\%. We thus assume that our predicted $\Bzdi$ is purely radial and contributes to the magnetic flux.
In \cite{vidotto2014dM_ZDI_winds}'s six wind models, the unsigned open magnetic flux at a large distance from the star, $\Phi_\textrm{open}$, is 30-56\% of the unsigned surface magnetic flux $\Phi_0$ from ZDI maps. We adopt the mean value of $f_\textrm{open,0}=$46\%.  We estimate the surface flux using our predicted large-scale field: $\Phi_0 = 4 \pi R_*^2 \Bzdi$.  The open flux is $\Phi_\textrm{open} = 4 \pi r^2 B(r) = f_\textrm{open,0} \Phi_0$, so:
\begin{equation}
    B(\tilde{r}) = f_\textrm{open,0} ~ \Bzdi ~ \tilde{r}^{-2} .
    \label{eq:B_open}
\end{equation}

\begin{figure}[ht!]
\epsscale{1.2}
\plotone{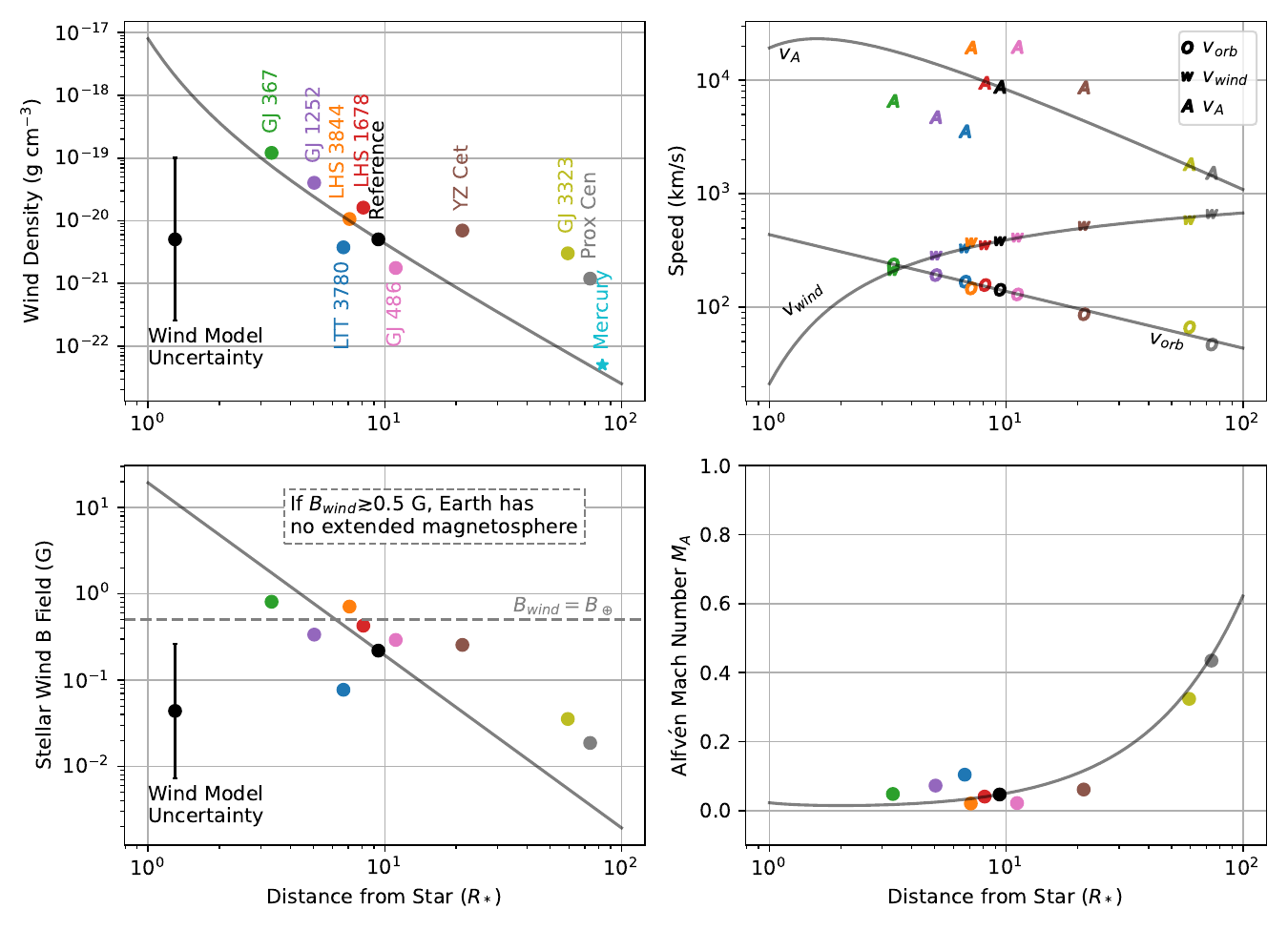}
\caption{Predicted wind parameters for our target systems (the 5 closest to the star) and comparison systems.  We show the radial dependence of wind parameters for a hypothetical reference system defined in Table~\ref{tab:targets} \textit{(gray lines)}.
\textit{(Top left)} Stellar wind density. Later-type M dwarfs have greater predicted mass loss flux and thus higher wind density relative to the reference line. The density of the solar wind at Mercury is shown for comparison. The error bar at left shows the scale of uncertainty for all systems due to inferring wind properties from rotation-activity relations (Section~\ref{sec:wind_uncertainty}).
\textit{(Top right)} Wind speed in stellar frame (w), planet orbital speed (o), and Alfv\'en speed (A).
\textit{(Bottom left)} Stellar wind magnetic field, modeled as an open radial field. Later-type M dwarfs have stronger predicted surface fields compared to the early-type reference system.  The dashed line shows a typical Earth field of 0.5~G; if the wind field exceeds this value, an Earth twin will not have an extended magnetosphere.
\textit{(Bottom right)} Alfv\'en Mach number of the wind relative to the planet. All of our targets can have sub-Alfv\'enic interactions ($M_A<1$) between the star and its innermost planet.
\label{fig:wind_model}}
\end{figure}

\subsection{Wind magnitude at planet: Velocity}

Solar/stellar winds accelerate significantly at the 3-10$R_*$ orbital distances of our target planets. We model this acceleration using a Parker wind \citep{parker1958solar_wind}, which has an analytical solution for wind speed \citep{cranmer2004parker} using the Lambert W function, available in \texttt{scipy} \citep{scipy2020NatMe..17..261V}.  This solution, derived from conservation of mass and momentum, assumes a steady-state isothermal wind with spherical symmetry (i.e., no angular expansion of open magnetic field lines), a better assumption for the equatorial slow solar wind than the polar fast wind. 
The 4 equations needed to implement this solution are reproduced here for ease of use:

\begin{align*}
\intertext{1) Isothermal sound speed assuming the wind consists of fully ionized hydrogen (atomic mass $m_H$, mean molecular weight $\mu=0.5$) at temperature $T$, where $T_6$ is in MK:}
    c_s &= \sqrt{ \frac{k T}{\mu m_H} } = \left( 128~\textrm{km\,s}^{-1} \right) \sqrt{T_6} \\
\intertext{2) Parker critical point $r_c$ where the wind exceeds the sound speed (in the rightmost expression, $M_*$ and $R_*$ are in solar units):}
    r_c &= \frac{G M_*}{2 c_s^2} = \left( 5.8 \, R_* \right) \frac{M_*}{R_* T_6} \\
\intertext{3-4) \citep[Eq.~15-16 in][]{cranmer2004parker} Solution for wind speed $u(r)$ using Lambert $W_0$ and $W_{-1}$ functions:}
    D(r) &= \left( \frac{r}{r_c} \right)^{-4} \exp{\left[ 4 \left( 1 - \frac{r_c}{r} \right) - 1 \right]} \\
    u &= \begin{cases}
        c_s \sqrt{ - \, W_0[-D(r)]} & r \leq r_c \\
        c_s \sqrt{ - \, W_{-1}[-D(r)]} & r \geq r_c
    \end{cases}
\end{align*}

Our targets' critical radius is $\sim$3$R_*$ compared to $\sim$6$\Rsun$ for the solar wind. This difference arises because our targets' hot coronae have higher thermal pressure gradients that cause rapid acceleration.

\subsection{Wind magnitude at planet: Density} \label{sec:rho_profile}

We determine wind density using mass continuity on open field lines:
\begin{equation}
    \rho(r) = \frac{\dot{M}}{f_{open}(r) ~ 4 \pi r^2 u}
\label{eq:wind_density}
\end{equation}
where $f_{open}(r)$ is the fraction of a sphere with radius $r$ surrounding the star that is pierced by open field lines. At the orbital distances of our targets, we assume $f_{open}(r)\approx 1$, which may slightly underestimate density near GJ~367~b at $3.4R_*$.

The Parker solution assumes the wind is purely radial, enabling an analytic solution, whereas a numerically-integrated Weber-Davis wind \citep{weberdavis1967wind} allows for azimuthal wind velocity caused by stellar rotation. We compared a radial Parker wind model for YZ~Cet to the Weber-Davis solution of \cite{pineda2023yzcet} for their base conditions, finding that the two are equivalent until $>$100$R_*$ since the star's slow rotation makes the azimuthal component negligible near the star ($u_r \gtrsim 100 u_\phi$).

Figure~\ref{fig:wind_model} shows our predicted stellar wind densities, which are 0.4-12$\times10^{-20}$~g\,cm$^{-3}$ near the planet in our 5 target systems.  Coincidentally, this mass density is comparable to that near Io \citep[$n\sim$400-2300~cm$^{-3}$ with 22 amu;][]{saur2013}.

\subsection{Uncertainties in Predicted Wind Conditions}
\label{sec:wind_uncertainty}

To interpret non-detections, we rely on the wind model for: 1) magnetic field strength near the stellar surface, which determines the maximum cyclotron frequency (Section~\ref{sec:ECM_freq}), and 2) wind magnetic field strength, 3) wind speed, and 4) wind density near the planet, which determine the Alfv\'en Mach number (Section~\ref{sec:M_A}) and the predicted SPI radio flux density (Section~\ref{sec:spi_model}).  Our predictions for all of these properties depend on stellar rotation period, which we estimate has $\sim$20\% uncertainty due to the varying quality of rotation period estimates for our targets (see references in Table~\ref{tab:targets}), but this is not the dominant source of uncertainty in the wind parameters due to large scatter in rotation-activity relationships. Throughout this work, to be conservative, we combine the orders-of-magnitude uncertainties in the wind model multiplicatively, rather than adding in quadrature, since the latter practice assumes independent variables.

\textit{1) Magnetic field strength near stellar surface.}  
75\% of a large sample of M~dwarfs follow the observational ZB relationship of \cite{reiners2022dM_ZB_correlation} to within $\times2$, so the uncertainty on our $\Bzb$ predictions (Equation~\ref{eq:B_ZB}) is $\times1.3$ (from $\Prot^{-1.25}$) $\times2$ (ZB scatter) $\approx \times 2.5$. In our Equation~\ref{eq:B_ZDI}, using $f_\textrm{ZDI}=$6\% and 14\% from \cite{reiners2009frac_BV_BI} introduces difficult-to-quantify uncertainty because these scale factors come from only 6 stars, all much more active than our targets. We take the uncertainty on $f_\textrm{ZDI}$ to be $\times 2$ (roughly the difference between the partially and fully-convective values), so that our estimates of large-scale field have uncertainty of order $\times5$.

\textit{2) Magnetic field strength near planet.}  Based on the range of values in \cite{vidotto2014dM_ZDI_winds}, we take the uncertainty in $f_\textrm{open,0}$ (in Equation~\ref{eq:B_open}) to be about 20\%, leading to a net uncertainty of $\approx \times 6$ in the circumplanetary stellar wind field strength (Figure~\ref{fig:wind_model}, lower left panel). Direct measurements of the stellar magnetic field are thus critical for interpreting SPI observations and identifying targets for SPI searches. While ZDI of all possible SPI targets would require prohibitive observing time, Zeeman broadening surveys could reduce the uncertainty in stellar wind magnetic field estimates to $\sim\times2.5$ instead of $\times 6$.

\textit{3) Wind speed near planet.} We estimate that our assumed wind temperature of 2~MK for slowly-rotating M~dwarfs has an uncertainty of $\times$2 (Section~\ref{sec:wind_temp}).  A factor of 2 change in wind temperature has a moderate effect ($\sim\times1.6$) on the wind speed at large distances, where a hotter wind reaches a higher terminal speed.  Temperature has a stronger effect in the wind acceleration region ($\times2-3$ at 3$R_*$) since a hotter wind accelerates more rapidly close to the star.  Thus, X-ray observations of coronal temperature, and/or theoretical estimates of wind heating, can improve wind speed and density predictions in the near-star regions relevant for sub-Alfv\'enic SPI. In an Alfv\'en-wave-driven model of TRAPPIST-1's stellar wind, \cite{reville2024dMwind_SPI} find that the possible range of slow and fast stellar wind speeds spans a factor of 2 depending on unknown parameters, consistent with our assumed wind speed uncertainty.

\textit{4) Wind density near planet.}  Our wind density predictions have two main uncertainties: wind temperature and mass loss rate. For mass loss rate, the uncertainty in rotation period and X-ray luminosity is a small effect compared to the observed scatter of $\pm\times10$ in the \cite{wood2021dM_winds} relationship, as noted for the SPI wind modeling of \cite{reville2024dMwind_SPI}. We note that one of our comparison systems, GJ~486, has an X-ray-based predicted mass loss rate of $0.05\dot{M}_\odot$, compared to a spindown-based law \citep{johnstone2015spindown_Mdot} predicting 1.4$\dot{M}_\odot$ \citep{pena-monino2025gj486_spi}. This wide discrepancy underscores the need for further observations and modeling of winds of slowly-rotating M dwarfs. We estimate a net uncertainty in density of $\times 20$ (Figure~\ref{fig:wind_model}, upper left panel): $\times 2$ from temperature and $\times 10$ in mass loss rate.

\section{Possible Causes of Non-Detections}\label{sec:nondetections}

There are multiple possible causes of a non-detection of radio SPI: 1) super-Alfv\'enic interaction, 2) timing and geometry, 3) frequency limits, and 4) flux density below detection limit.  In this section, we rule out \#1 as unlikely and consider options \#2 and \#3, while we discuss \#4 in Section~\ref{sec:spi_model}.

\subsection{Super-Alfv\'enic interaction}
\label{sec:M_A}

SPI cannot power stellar radio emission if the interaction is super-Alfv\'enic, i.e.~the stellar wind speed relative to the planet exceeds the local Alfv\'en speed. We used our wind model to calculate the Alfv\'en speed for our targets, predicting $M_A\lesssim0.1$ for our 5 targets (Figure~\ref{fig:wind_model}). We also predict that all of our comparison sample are sub-Alfv\'enic, including Prox Cen~b with $M_A\sim0.5$, but in a more sophisticated 3D model of Prox Cen's wind, \cite{kavanagh2021} find that the wind should become super-Alfv\'enic at $\sim 40-70 R_*$ for a wide range of stellar mass loss rates. The difference may arise because of our use of magnetic field correlations instead of observed quantities. Our 5 primary targets are so close to the star that even considerably lower stellar magnetic fields than predicted would not make them super-Alfv\'enic.

\subsection{Timing \& geometry}

The timing of observations, and the geometry of the system, could cause non-detections due to either angular beaming or luminosity changes due to stellar wind variability.  \cite{kavanagh2023geometry} predict that angular beaming will cause transiting systems to be visible 4\% of the time, and only 49\% of transiting systems will ever direct their emission towards Earth. As a caveat, this study assumes a uniform distribution for the unknown stellar inclination and magnetic obliquity whereas if these are preferentially nearly aligned with the orbital axis the detection rate may be higher. To account for the 4\% problem, we observed almost a full orbital period for all of our targets (Table~\ref{tab:obs}).  ZDI observations could enable tailored simulations to help address whether individual systems are in the ever-observable 49\%.

\subsection{Footpoint conditions \& observing frequency}
\label{sec:ECM_freq}

The magnetic flux tube connecting planet and star ends at a stellar footpoint, producing radio emission at the cyclotron frequency at a range of heights along the flux tube, up to a maximum surface cyclotron frequency if not absorbed by the stellar wind. \cite{pena-monino2025gj486_spi} and \cite{pena-monino2025radioSPI_code} considered the possibility of free-free absorption of an ECM signal on GJ~367 and YZ~Cet, respectively.  They found that for coronal temperatures around 2~MK, absorption is negligible for mass loss rates $<10\dot{M}_\odot$ (GJ~486) and $<2\dot{M}_\odot$ (YZ~Cet), well above our targets' predicted mass loss rates (Table~\ref{tab:wind}). However, an emission cutoff below our observed frequency band may explain the observed non-detections.

M~dwarfs, including slow rotators such as YZ~Cet \citep{pineda2023yzcet} and GJ~3323 \citep{ortiz2024gj3323_exoradio}, are known to emit polarized emission (likely ECM) at GHz frequencies, and polarized bursts on active M dwarfs such as AD~Leo may connect to large-scale magnetic structures \citep{zarka2025adleo}. Motivated by such past detections of GHz coherent bursts on M~dwarfs, we observed at 1-4~GHz in order to tap into the high sensitivity of GHz radio facilities.  However, GHz bursts could also originate from small-scale closed magnetic loops that do not connect magnetically to the stellar wind, as seen for solar ECM bursts by \cite{yu2024solar_ECM}.

Table~\ref{tab:wind} predicts magnetic fields for our targets, with average surface fields (including small-scale flux) of 130-560~G, and large-scale fields of 8-80~G, with an estimated uncertainty of order $\times5$.  These predicted fields are lower than those of our comparison sample because our current targets are mostly earlier spectral type. If the planet field lines connect to weakly magnetized coronal holes, similar to the fast solar wind, then the average large-scale field may provide a good estimate of the ECM cutoff frequency, which would be of order 20-250~MHz for our targets.  However, if the planet field lines connect to coronal streamers, like the slow solar wind, then the footpoints may land near active regions with enhanced field and thus reach cyclotron frequencies of 0.4-1.6~GHz.  With the significant uncertainty in the predicted stellar magnetic fields, it is plausible that our targets may have SPI-driven ECM above 1~GHz, but the ECM cutoff frequency also provides a reasonable explanation for all of the non-detections. Stellar magnetic field observations of close-in exoplanet hosts can help select appropriate observing frequencies to optimize sensitivity in the search for radio SPI.

\vspace{1em}

While geometry and observing frequency are plausible explanations for our non-detections, another possible explanation is flux density too low to detect, a case that would give constraints on exoplanet magnetospheric size and field strength, as we discuss in the following sections.

\section{Upper limits on exoplanet magnetosphere}
\label{sec:spi_model}

Our goal is to translate between observed radio flux density (upper limits) and exoplanet magnetic field strength, which determines the interaction cross-section and thus the SPI power. First, we generate ``minimum'' SPI power predictions for a planet-sized magnetosphere. Then, we compare these predictions to observations to place upper limits on the target planets' magnetospheric size and field strength.

\subsection{Alfv\'en wing model for SPI power}

\cite{callingham2024radio_stars_review} review various models for predicting SPI power. Here we adopt the widely-used Alfv\'en wing model \citep[][S13]{saur2013}, which has been benchmarked against solar-system observations and numerical simulations. We wrote independent code for this project, but we note that the Alfv\'en-wing model has recently become available in the SPI modeling code SIRIO \citep{pena-monino2025radioSPI_code}.

In the Alfv\'en wing model, the SPI power is the integrated Poynting flux leaving the planet and traveling back towards the star in one of two Alfv\'en wings, since in an open stellar field the other wing does not connect to the star.  \cite{saur2013}'s Equation~55 for SPI power, which we denote $\Pspi$, is valid for low Alfv\'en Mach number $M_A$.  The underlying approximation that $(1-1/M_A^2)^{-1/2} \approx M_A$ \citep{callingham2024radio_stars_review} is accurate to within 20\% for $M_A<0.8$. \cite{pineda2018} reframe this equation in cgs to show the dependence of $M_A$ on wind density:
\begin{equation}
    \Pspi = \frac{1}{2} \abar^{2} \Raw^{2} v^2 B \sin^{2} \theta \sqrt{4\pi \rho} \; .
    \label{eq:saur}
\end{equation}
The parameters in this model are: a dimensionless interaction strength $\abar$ (Section~\ref{sec:abar}), the cross-sectional radius $\Raw$ of the Alfv\'en wings (Section~\ref{sec:Raw}), the stellar wind's velocity in the planet's rest frame at speed $v$ at angle $\theta$ relative to $\vec{B}$ (Section~\ref{sec:theta}), and the unperturbed stellar wind magnetic field $\vec{B}$ and mass density $\rho$ at the planet's location (Section~\ref{sec:wind}).  The unperturbed stellar wind properties are the values that would occur if the planet was not present.

\subsection{Interaction strength $\abar$} \label{sec:abar}

The interaction strength parameter $\abar$ can vary from 0 (no interaction) to 1 (strong interaction), quantifying the satellite's ability to halt oncoming plasma flow through electromagnetic interaction.  Strong interaction ($\abar\approx 1$) occurs in two cases: objects with strong extended magnetospheres such as Ganymede, and unmagnetized objects with an electrically-conducting ionosphere such as Io.  In contrast, unmagnetized atmosphere-less objects such as Tethys and Dione can have low interaction strength \citep[$\abar \sim 10^{-2}$;][]{saur2013}.

We assume $\abar=1$ in our calculations.  Since most of our radio upper limits correspond to extended magnetospheres (Section~\ref{sec:UL_Rmag}), this assumption should be appropriate. However, if the resulting constraint on the Alfv\'en wing radius were less than the planet size, the upper limit could reflect a low interaction strength due to the lack of an atmosphere, which is likely for our systems (Section~\ref{sec:targets}).

\subsection{Radius of Alfv\'en wing} \label{sec:Raw}

The cross-sectional radius of the Alfv\'en wing, $\Raw$, is measured far enough from the planet that the field lines connecting the star and planet have become parallel. Here, $\Raw$ is the equivalent of $R$ or $\Reff$ in S13, or $R_o$ in \cite{pineda2018}.  Determining $\Raw$ depends on whether the planet has a closed magnetosphere that extends above its surface.

\textit{Extended magnetosphere: $\Raw = \sqrt{2} \Rmag$.}  If the planet has a strong enough magnetic field, it will form a closed magnetosphere with radius $\Rmag$ (denoted $R_\textrm{obst}$ in S13) in pressure balance with the stellar wind.  Depending on the relative orientation $\Theta_M$ of the stellar wind magnetic field and planet's equatorial field, the Alfv\'en wing may expand wider than the closed magnetosphere, or in the relatively rare case of anti-aligned fields, no stellar field lines connect to the planet and the Alfv\'en wing does not exist.  Thus,
\begin{equation}
    \Raw = \fgeo \Rmag
\label{eq:f_geo}
\end{equation}
where $\fgeo \approx \sqrt{3 \cos ( \Theta_M / 2)}$ ranges from 0 to $\sqrt{3}$ depending on orientation (S13).  Since $\Raw^2$ appears in SPI power, we perform a spherical average of $\fgeo^2$ across all $\Theta_M$ to obtain $\langle \fgeo \rangle=\sqrt{2}$.  This is close to the value obtained for $\theta_M=90\degr$, which may be a preferred orientation in our systems, since the tidally-locked planetary rotation axis should be nearly perpendicular to the radial stellar wind field.  Thus, we adopt $\Raw = \sqrt{2} \Rmag$ for planets with extended magnetospheres.

\textit{Induced magnetosphere: $\Raw = R_p$.}  If the planet is unmagnetized or its field is too weak to form an extended magnetosphere, then if it has an ionosphere it can form an induced magnetosphere.  S13 adopts $\Raw = 1.3 R_p$ in that case for unmagnetized solar system moons such as Io.  However, for the gravity of a terrestrial planet rather than a moon, the ionosphere may be thinner than Io's, so we adopt the simple assumption that $\Raw = R_p$ for weakly magnetized planets.

\subsection{Stellar wind direction: simplifying $v \sin \theta$} \label{sec:theta}
 
\begin{figure}[ht!]
\epsscale{1.2}
\plotone{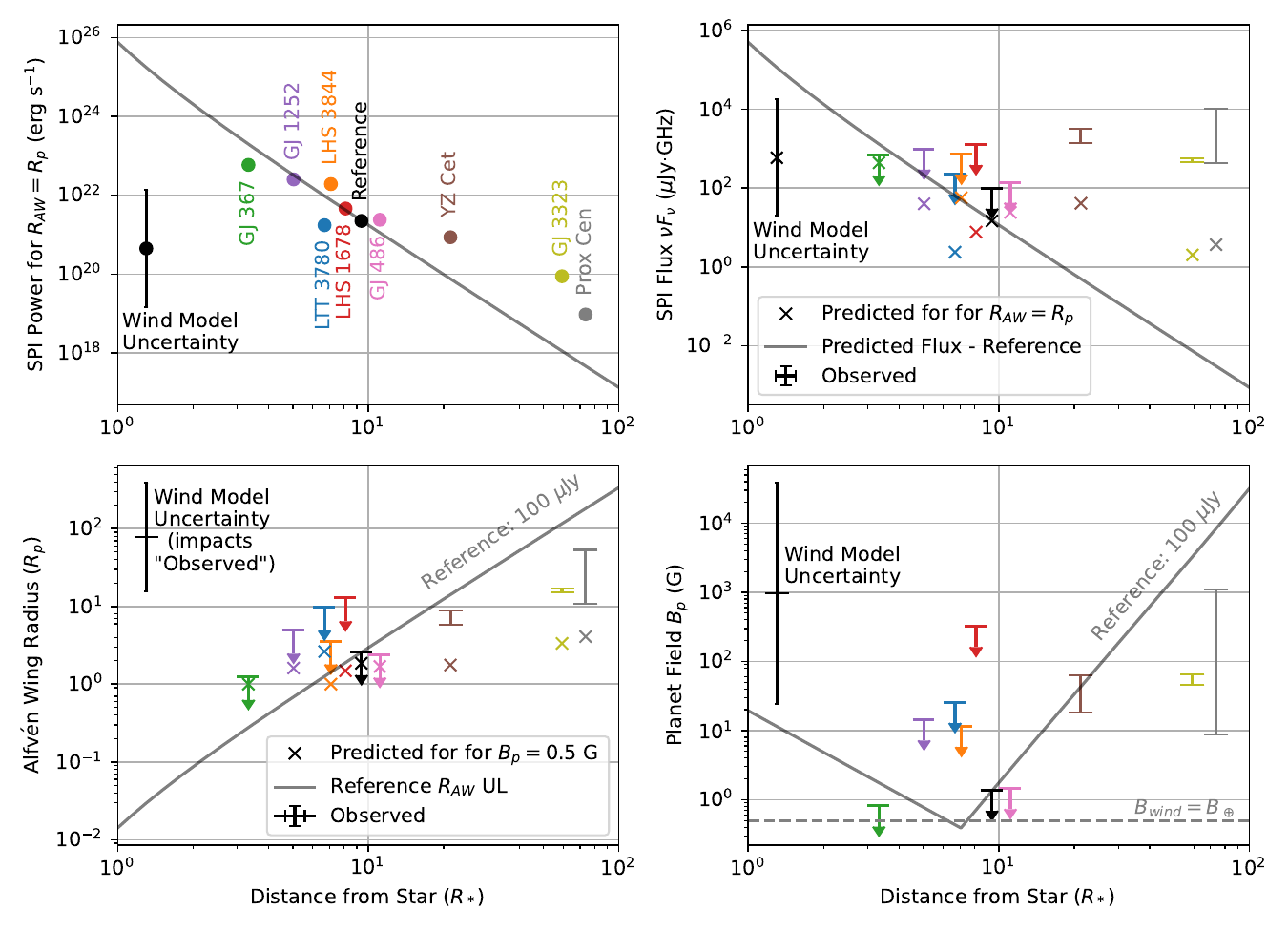}
\caption{
\textit{(Top left)} Predicted SPI power for an Alfv\'en wing the width of the planet.  For a given system, the predicted power declines sharply with orbital distance, but the later spectral types of our comparison sample somewhat offset their greater orbital distances. The error bar at left shows the scale of uncertainty for all systems due to inferring wind properties from rotation-activity relations (Section~\ref{sec:spi_uncertainty}).
\textit{(Top right)} Observed radio upper limits or detections for our targets and comparison sample, compared to predicted SPI flux \textit{($\times$ marks)} for an Alfv\'en wing the width of the planet. We apply the estimate $\Delta \nu = \nu_{cen}$ so $\nu F_\nu$ is an estimate of spectrally-integrated flux. For comparison systems YZ~cet and Prox~Cen, the error bars denote the range of observed coherent burst flux densities in different epochs, whereas for GJ 3323 the error bars show the uncertainty on a single-epoch detection.  \textit{(Bottom left)} The Alfv\'en wing radius needed to match the observational constraints, compared to its predicted size for an Earth-like planet field of 0.5~G.  We also show how the sensitivity to Alfv\'en wing size varies with orbital distance, for a radio sensitivity of 100\ujy ~on the reference system.  Closer-in orbits offer sensitivity to smaller Alfv\'en wings, and thus to weaker exoplanet magnetic fields. The error bar at left shows the wind model uncertainty for all systems that enters when converting flux density to Alfv\'en wing radius, which is not included in the $3\sigma$ upper limits.
\textit{(Bottom right)} Exoplanet magnetic dipole field strength consistent with observed flux densities or upper limits.  The reference line bends near the star at the point where the detectable $\Raw=R_p$. In this reference system, to the left of this bend, the upper limit shown is the largest $B_p$ that would not create an extended magnetosphere around the planet, which is $B_p = B_\textrm{wind}$.
\label{fig:spi_model}}
\end{figure}

The SPI power equation (Equation~\ref{eq:saur}) depends on the angle $\theta$ between the stellar wind magnetic field $\vec{B}$ and its flow direction $\vec{v}$ in the planet's rest frame.  We assume the planets are in circular orbits (Section~\ref{sec:targets}). For our radial wind model (Section~\ref{sec:wind}), the wind velocity and stellar magnetic field are both radial in the star's rest frame.  In the planet frame, the magnetic field remains radial away from the star, but the stellar wind speed $\vec{v}$ has an added azimuthal component of the orbital velocity. Taking $v \sin \theta$ gives only the component of $\vec{v}$ perpendicular to the radial $\vec{B}$, so only the azimuthal component, so $v \sin \theta \approx \vorb$. To refine this assumption, in Appendix~\ref{app:v_sin_theta}, we show that for a Weber-Davis wind appropriate for a rotating star, if stellar rotation and orbital axes are aligned, the expression instead simplifies to $v \sin \theta = \vorb (B_r/B) (1-\Porb/\Prot) \approx \vorb (1-\Porb/\Prot)$ for our slowly-rotating sub-Alfv\'enic systems. We implement this latter approximation in our SPI power predictions in Figure~\ref{fig:spi_model}.

\subsection{Predicted SPI power for unmagnetized planet}
\label{sec:Pspi_unmag}

The top left panel in Figure~\ref{fig:spi_model} shows our predictions for the SPI power transferred from near the planet back towards the star, for the unmagnetized case where the Alfv\'en wing is the width of the planet.  The radial dependence of this power with orbital distance $a$ is:
\begin{equation}
\Pspi \propto \vorb^2 B \sqrt{\rho}
      \propto \left( \frac{1}{\sqrt{a}} \right)^2
      \frac{1}{a^2} \sqrt{\frac{1}{a^2 \vwind}}
      \propto \frac{1}{a^4 \sqrt{\vwind}}.
\end{equation}
Accounting for wind acceleration, $\Pspi \propto a^{-5}$ to $a^{-4}$ at 3-100$R_*$.  This strong dependence is somewhat reduced for magnetized planets.  For a given planetary field strength, more distant planets will carve out a larger magnetosphere in balance with the decreased stellar wind magnetic pressure, reflected in $\Rmag \propto a^{2/3}$ (Section~\ref{sec:Rmag_to_B}), so $\Pspi \propto a^{-3.7}$ to $a^{-2.7}$.  Close-in planets dramatically improve the prospects for detecting SPI, but better stellar conditions (late type stars close to Earth with strong stellar fields and mass loss flux) can outweigh this advantage.

\subsection{Predicted SPI radio flux for unmagnetized planet} \label{sec:predict_flux}

For the case of an unmagnetized planet, we next translate our predictions of SPI power into predictions of radio flux.  We predict the spectrally-integrated radio flux $F_R$ since it is  independent of the unknown bandwidth $\Delta \nu$, which means these predictions can be easily adapted to the observing frequencies of various radio telescopes. The predicted radio flux is:
\begin{equation}
    F_R = F_\nu \Delta \nu = \frac{\eta \Pspi}{\Omega d^2}
\end{equation}
where $\eta$ is the efficiency of converting SPI power into ECM radio luminosity, $\Omega$ is the solid angle of the beamed ECM radiation, and $d$ is the distance to Earth.  We adopt the Solar System-grounded assumptions used in e.g., \cite{pineda2023yzcet} of $\eta=0.01$ and $\Omega=0.16$~ster.  Figure~\ref{fig:spi_model} (bottom left) shows our predicted fluxes for the unmagnetized planet case.  These range from 4-440~$\mu$Jy$\cdot$GHz, where GJ~367's prediction is the highest due to strong stellar wind conditions at $a/R_*\sim3$. 

In Table~\ref{tab:targets}, we give flux density upper limits for an integration time of 5 minutes.  For comparison system GJ~486, we scaled from the dynamic-spectrum Stokes V sensitivity in \cite{pena-monino2025gj486_spi} up to an integration time of 5 minutes and the full bandwidth.  While M dwarf coherent bursts show a wide range of durations in the literature, we chose a 5-minute duration to be order-of-magnitude consistent with the time for a beam thickness of a few degrees (beam width plus planet magnetosphere's angular extent) to sweep across the line of sight during a $\sim$1-day orbit.

We converted observational radio upper limits and detected radio flux densities for some comparison systems (Table~\ref{tab:targets}) to spectrally-integrated flux (Figure~\ref{fig:spi_model} upper right) using the common assumption for radio SPI observations that $\Delta \nu = \nu_\textrm{cen}$, where $\nu_\textrm{cen}$ is the center frequency of the observed band.  The observational constraints lie above the predictions for all systems except GJ~376, where the upper limit is close to the observed value. This difference implies that most systems require an extended magnetosphere to be detectable in our observations.

\subsection{Upper limits on magnetospheric radius} \label{sec:UL_Rmag}

We can then solve for Alfv\'en wing radius:
\begin{equation}
   \frac{\Raw}{R_p} = \sqrt{ \frac{\textrm{observed flux}}{\textrm{predicted flux for $\Raw=R_p$}} }
\end{equation}
The results are shown in Figure~\ref{fig:spi_model}, where our 5 radio upper limits correspond to Alfv\'en wing sizes of 1.3$R_p$ (GJ~367) up to 12.9$R_p$ (LHS~1678).  In the comparison sample, GJ~486's radio upper limit leads to $\Raw<2.4 R_p$, whereas the three systems with radio burst detections require $\Raw\sim8-40R_p$ to explain the range of observed flux densities.  In the same figure, we also show theoretical predictions of the Alfv\'en wing radius (see next section) for an Earth-like planetary field of 0.5~G, which range from 1-2.4\,$R_p$ for our 5 targets and 1.7-4.1\,$R_p$ for the comparison sample since its more distant planets experience lower wind pressures.  In our sample, a prediction of $1R_p$ occurs when the stellar wind field is stronger than 0.5~G (LHS~3844 and GJ~367).  For comparison, Earth's magnetosphere has a dayside extent of 6-10$R_E$ \citep{pulkkinen2007earth_magnetopause}, larger than expected for our targets due to the lower wind pressures at Earth. 

Next, we use $\Raw = \sqrt{2} \Rmag$ (Section~\ref{sec:Raw}) to estimate the magnetospheric sizes of our targets, obtaining upper limits of $\Rmag<$1-9$R_p$.  If $\Raw<\sqrt{2}R_p$, we impose $\Rmag<1R_p$.  

\subsection{Upper limits on exoplanet magnetic field}
\label{sec:Rmag_to_B}

We follow S13 in assuming that at the magnetospheric boundary, the exoplanet's magnetic pressure balances the stellar wind pressure.
We apply our stellar wind model (Section~\ref{sec:wind}) to include magnetic, dynamic, and thermal pressure; Figure~\ref{fig:spi_model} (lower right) shows the resulting upper limits on a dipolar $B_p$, which range from 0.8~G (GJ~367) to 320~G (LHS~1678).

While our calculations include all 3 types of stellar wind pressure, we note that thermal pressure is negligible for all targets, and dynamic pressure nearly so, since $P_\textrm{dyn} = 2 M_A^2 P_{mag}$ and all of our targets are sub-Alfv\'enic.  Thus, our results effectively simplify to the planetary and stellar magnetic pressure being equal at the magnetospheric boundary. We assume the exoplanet field $B_p$ is dipolar so:
\begin{equation}
B_\textrm{wind} \approx B_p \, (\Rmag/R_p)^{-3}.
\label{eq:Rmag}
\end{equation}

GJ~367 highlights an interesting limitation on sub-Alfv\'enic SPI measurements of planetary magnetic field for close-in planets.  If the stellar wind field is stronger than the planet's field, then the planet has no extended magnetosphere.  Thus, since our radio upper limits for GJ~367 were consistent with $\Rmag = R_p$, our upper limit on GJ~367~b's planetary magnetic field is essentially the predicted strength of the stellar wind field at that orbital distance (also seen in Figure~\ref{fig:wind_model}, lower left).  This situation is also reflected in the curve showing the reference system's hypothetical 100-$\mu$Jy upper limit on $B_p$ as a function of orbital distance, which bends upwards left of $<\sim7R_*$ to follow $B_p<=B_\textrm{wind}$.  For extremely close-in terrestrial planets, they are unlikely to have strong enough magnetic fields to carve out extended magnetospheres, and thus sub-Alfv\'enic SPI can only place upper limits on their magnetic field strengths even if there were a detection.

\subsection{Effects of uncertainty in modeling SPI}
\label{sec:spi_uncertainty}

The level of uncertainty in the SPI model depends on whether the interaction occurs at the ionosphere or in the extended magnetosphere.  For the the assumption of strong ionospheric interaction ($\bar{\alpha}=1$ and $\Raw = R_p$), used in the baseline predictions in Figure~\ref{fig:spi_model} (upper panels), the uncertainty in predicted SPI power and flux density is $\approx \times30$ due to $\sqrt{\rho}\sim\sqrt{\times20}\sim\times5$  and $B_\textrm{wind} \sim \times6$ (Section~\ref{sec:wind_uncertainty}). The uncertainty in $\Pspi$ translates into an uncertainty in inferred Alfv\'en Wing radius of $\times \sqrt{30} \approx \times 5$ (Figure~\ref{fig:spi_model}, lower left panel).

We estimate the uncertainty in $B_p$ assuming an extended magnetosphere. For such magnetospheric interaction, solving for upper limits on exoplanet magnetic field depends on wind conditions following $B_p \propto B_\textrm{wind}^{-1/2} ~ \rho^{-3/4} ~ \fgeo^{-3}$ (Equations~\ref{eq:saur}, \ref{eq:f_geo} and \ref{eq:Rmag}). The extended-magnetosphere case thus reduces the impact of the uncertainty in stellar field, but introduces a strong dependence on the unknown relative orientation of the wind and planetary magnetic field $\Theta_M$ through the geometric factor $\fgeo$ (Section~\ref{sec:Raw}). If we assume a random orientation of $\Theta_M$ on a sphere, then there is a 68\% probability that $\Theta_M$ lies in the range 47$^\circ$-133$^\circ$ and thus that $\fgeo$ lies between 1.1-1.7, yielding a fractional uncertainty on $\fgeo^{-3}$ of 70\%.  Hence, the combined effects of wind density, field, and geometry yield an uncertainty in $B_p$, beyond the flux density uncertainty factored into the 3$\sigma$ upper limits, of $\times 20^{3/4} \times1.7 \times\sqrt{6} \approx \times 40$ (Figure~\ref{fig:spi_model}, lower right panel).

The uncertainty introduced by unknown model parameters, and by choices in modeling approach, becomes evident when comparing our results to literature models of our comparison systems. For example, our inferred exoplanet field to explain YZ Cet's observed radio bursts is at least 20~G, higher than the $\sim$10~G found by \cite{pineda2023yzcet} with their analogous wind Model~B.  The discrepancy arises primarily because of different assumptions for $\fgeo$ (Section~\ref{sec:Raw}): we assume that the Alfv\'en wing is $\sqrt{2}$ times larger than the closed magnetosphere, reflecting perpendicular stellar wind and planet fields, whereas \cite{pineda2023yzcet} assumed the maximum (field-aligned) value of $\sqrt{3}$, and $B_p \propto \fgeo^{-3} \sim \sqrt{2/3}^{-3} \sim 1.8$.

For another comparison to literature SPI models, we ran our models with a range of wind temperatures (Section~\ref{sec:wind_uncertainty}), inspired by the parameter exploration of \cite{pena-monino2025gj486_spi}(PM25) for GJ~486.  PM25 predicted that SPI flux density increases with wind temperature (almost proportionally), likely due to faster wind flow past the planet. In contrast, we found that a factor of 2 increase in wind temperature decreased SPI power by $\times1.6-1.25$ at 3-100$R_*$. This relatively weak dependence occurs in our model because the uncertainty in $\vwind$ due to unknown wind temperature does not significantly affect $v \sin \theta$ (which $\approx \vorb$ for close-in planets around slow rotators, Section~\ref{sec:theta}) nor the planet's magnetospheric size since wind magnetic pressure dominates over dynamic pressure (Section~\ref{sec:Rmag_to_B}), and thus the only dependence of SPI power on wind speed is through $\sqrt{\rho}\sim1/\sqrt{\vwind}$, where higher temperatures lead to faster $\vwind$ acceleration. These contrasting results point to different choices in modeling approach, potentially in addressing the relative orientation $\theta$ between wind velocity and magnetic field in the planet frame.

With the many uncertainties involved, our calculations demonstrate the plausibility of detectable radio SPI from these systems, but observations and theoretical modeling of the magnetized stellar wind can refine predictions, as recently undertaken for YZ~Cet by \cite{pineda2026yzcet_zdi}.

\section{Conclusions}
\label{sec:conclusions}

We conducted GHz radio observations of 5 slowly-rotating M dwarfs with close-in exoplanets to search for stellar radio bursts induced by sub-Alfv\'enic SPI: LTT~3780 at 2-4~GHz with the VLA, and LHS~3844, GJ~367, LHS~1678, and GJ~1252 at 1.1-3.1~GHz with ATCA.  We observed each target for most or all of an orbital period.  In time series with 5-minute time binning, we did not detect any radio bursts in Stokes I or V, with Stokes V $3\sigma$ upper limits of 75~\ujy\ (VLA) and 300-600~\ujy\ (ATCA).
    
In images of the full observation, we detected quiescent Stokes~I emission from LHS~3844 at 63$\pm$12~\ujy\ and tentatively detected quiescent Stokes~V emission from LHS~1678 at -70$\pm$9~\ujy, although poor source coherence in the latter case makes the detection tentative.  Lack of temporal variability means that neither of these signals are consistent with the minutes-to-hours long bursts expected as angularly-beamed emission from a planet-connected flux tube sweeps past the viewer, but they could be consistent with active regions on the slowly-rotating stellar surface.  LHS~1678's early-M type and low photometric variability \citep{kar2024AJphot_var} make its polarized radio emission particularly surprising. Quiescent emission from exoplanetary hosts provides background illumination that can enable exoplanet radio transit observations, if these targets are confirmed as quiescent emitters in multiple epochs.  The radio transit profile should be modified by exoplanet magnetospheres, but for uniform stellar emission, a small fractional transit depth would require next-generation radio facilities \citep{hazra2022radio_transit}.  However, with luck, current facilities could detect high transit depths due to the concentration of stellar radio emission in small active regions \citep{pope2019radio_transit}.

To interpret the non-detection of SPI radio bursts, we constructed a stellar wind model for our targets (Section~\ref{sec:wind}). We used a radial 1D isothermal Parker wind, and a radial open magnetic field whose strength is scaled down to account for most of the surface field not connecting to open field lines.  Since our targets did not have magnetic or stellar wind measurements, we adopt prescriptions for predicting these values from literature rotation-based correlations. While the uncertainties in mass loss rate and stellar magnetic field lead to $\sim\times30$ uncertainty in the predicted SPI radio flux density, these rotation-based wind predictions enable population studies and strategic target selection for cases where it is not feasible to perform detailed magnetic/wind observations and modeling for every target.

The radio SPI non-detection may be caused by unlucky geometry with beamed emission never directed towards Earth, or an observing frequency above the maximum cyclotron frequency on the star-planet magnetic flux tube, or an SPI flux density too low to detect.  We predict that all of our targets are sub-Alfv\'enic and thus can transfer energy to radiate near the star; the predicted stellar large-scale surface fields correspond to cyclotron frequencies of 20-250~MHz, but small-scale surface fields may boost local cyclotron frequencies into the GHz range.

If the non-detections are due to low flux density, our stellar wind model implies upper limits (Section~\ref{sec:spi_model}) on planet magnetospheric size of 1-9 planet radii, and planet magnetic fields of $<$0.8~G for GJ~367 (the closest-in planet at $a/R_*=3.3$) to $<320$~G for LHS~1678 (the most distant at $a/R_*=8$).  Notably, GJ~367~b's upper limit corresponds to the circumplanetary stellar wind magnetic field strength; since planetary fields lower than this would not create an extended planetary magnetosphere, even SPI detections cannot measure planetary fields weaker than the stellar wind field.

While there are many uncertainties in SPI flux density predictions, our wind model enables us to compare the relative predicted SPI power and the informativeness of upper limits across exoplanetary systems.  We compare our 5 targets to 4 other slowly-rotating M dwarfs with radio SPI candidate detections (YZ Cet, Prox Cen, GJ 3323) and upper limits (GJ~486). Next to GJ~367, we find that GJ~486 provides the tightest constraint on exoplanet magnetic field, $<$1.5~G for our wind model.  Curiously, the 3 radio-detected comparison systems have the largest orbital distances ($a/R_*\sim$20-70) and thus relatively low predicted flux densities, although the strong stellar magnetic fields predicted for their late spectral type somewhat offset the effects of orbital distance.  This disparity, of radio detections in systems with low predicted SPI power, could easily occur if the detected radio bursts are due to non-planetary stellar activity. Or, if the bursts are indeed SPI, it indicates unusually high stellar wind magnetic fields or densities that increase the SPI power or radio emission efficiency, or that later-type M dwarfs are more likely to emit up to GHz cyclotron frequencies on the star-planet flux tube, compared to the early-to-mid M slow rotators targeted in this survey.  Our non-detections of targets with relatively high predicted SPI power collectively favor an approach of observing at sub-GHz frequencies to search for SPI on inactive early-to-mid M dwarfs.

\section{Acknowledgments}


JV thanks Aline Vidotto for helpful advice about stellar winds. This material is based upon work supported by the National Science Foundation under Grant No.\ AST-2150703 and AST-2310589 (JV, LG, EH, AW, AA) and AST-2108985 (JSP).  The National Radio Astronomy Observatory is a facility of the National Science Foundation operated under cooperative agreement by Associated Universities, Inc. Support for this work was provided by the NSF through award SOSP 19B-005 from the NRAO (CR).  The Australia Telescope Compact Array is part of the Australia Telescope National Facility (grid.421683.a) which is funded by the Australian Government for operation as a National Facility managed by CSIRO. We acknowledge the Gomeroi people as the traditional owners of the Observatory site.
This work has made use of data from the European Space Agency (ESA) mission
{\it Gaia} (\url{https://www.cosmos.esa.int/gaia}), processed by the {\it Gaia}
Data Processing and Analysis Consortium (DPAC,
\url{https://www.cosmos.esa.int/web/gaia/dpac/consortium}). Funding for the DPAC has been provided by national institutions, in particular the institutions participating in the {\it Gaia} Multilateral Agreement. This research has made use of the SIMBAD database, operated at CDS, Strasbourg, France. This research made use of Astropy (\url{http://www.astropy.org}), a community-developed core Python package for Astronomy.

\software{Astropy \citep{astropy:2013, astropy:2018, astropy:2022}, Scipy \citep{scipy2020NatMe..17..261V}}

\clearpage
\bibliography{references}{}
\bibliographystyle{aasjournal}


\appendix

\section{Images} \label{app:images}

\begin{figure}[ht!]
\gridline{\fig{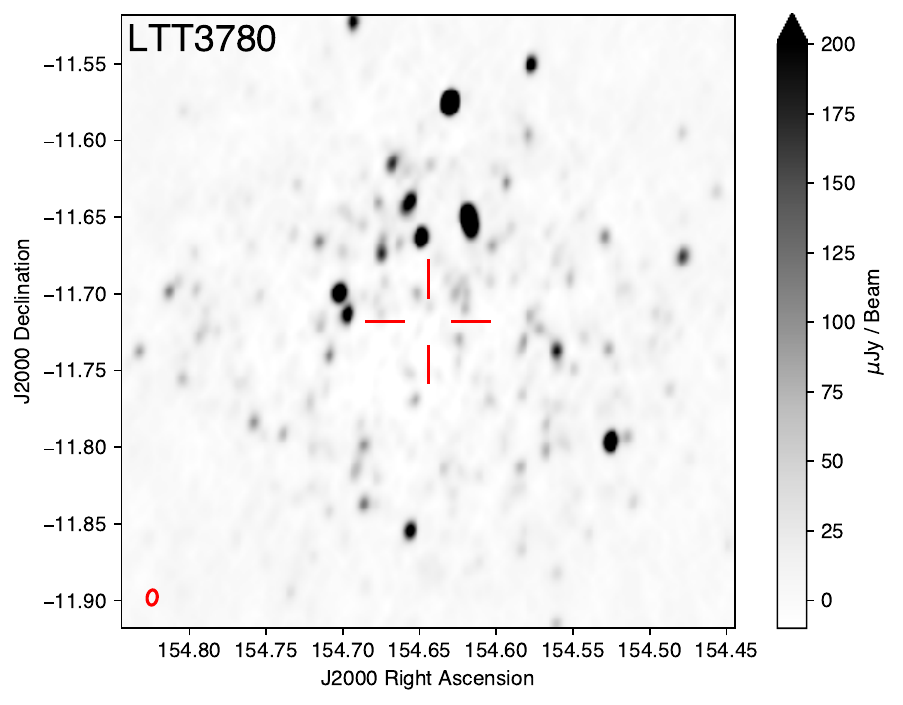}{0.45\textwidth}{}
          \fig{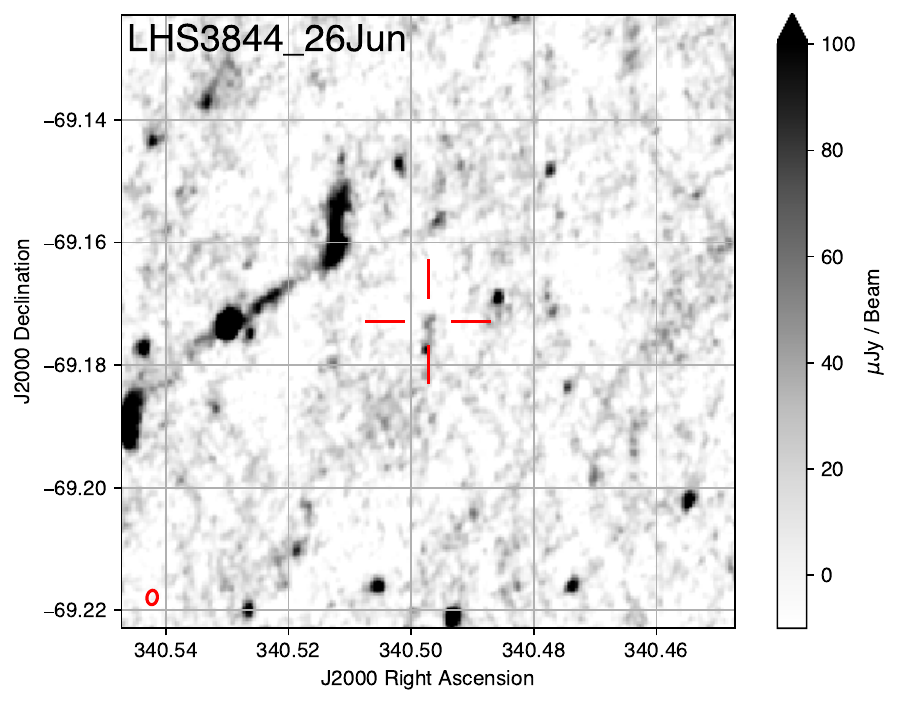}{0.45\textwidth}{}}
\vspace{-2em}
\gridline{\fig{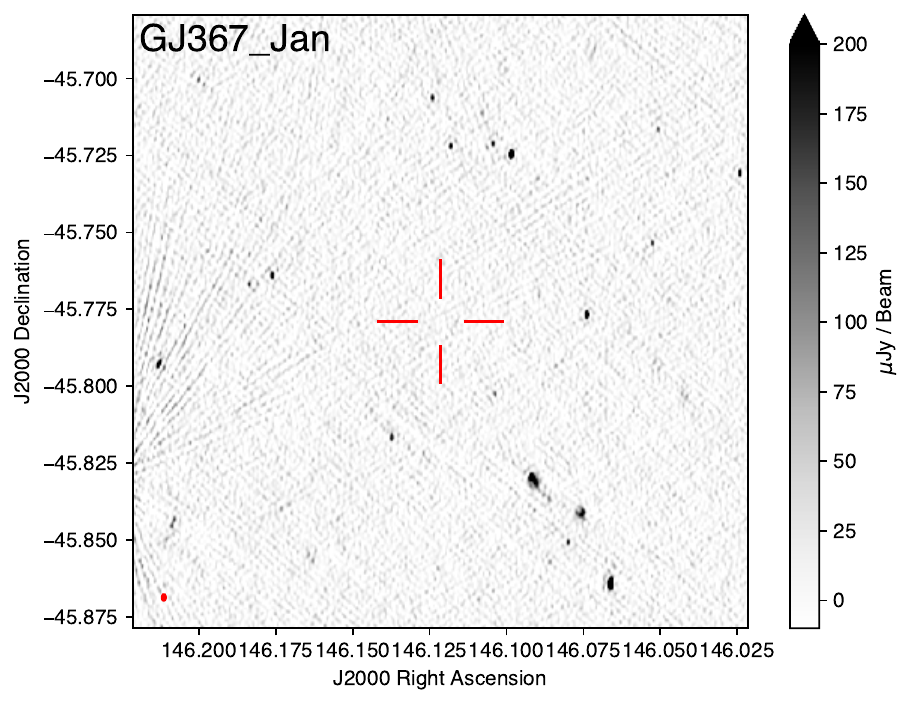}{0.45\textwidth}{}
          \fig{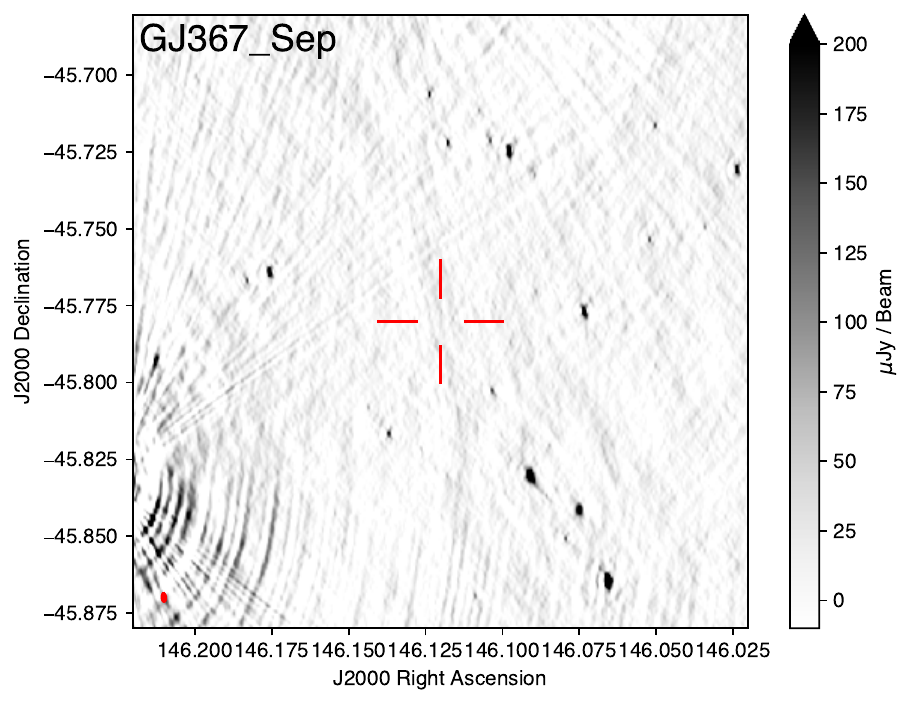}{0.45\textwidth}{}}
\vspace{-2em}
\gridline{\fig{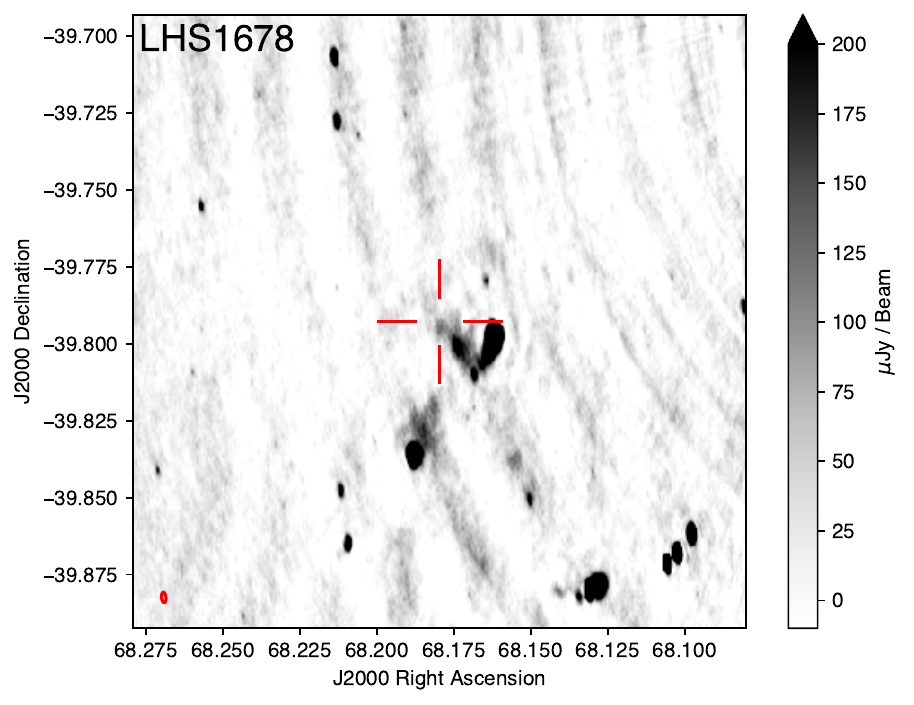}{0.45\textwidth}{}
          \fig{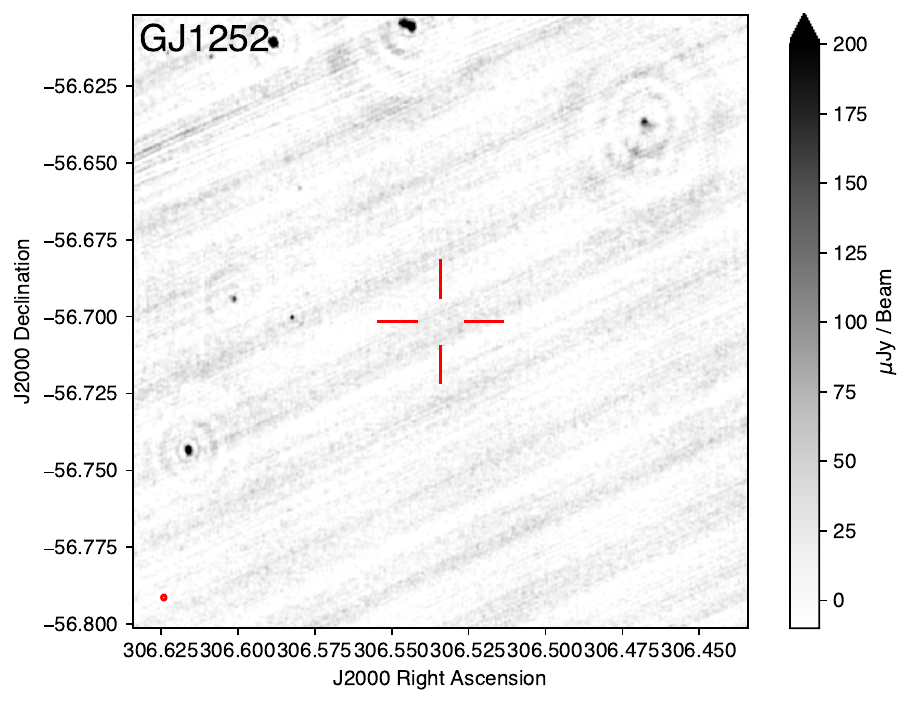}{0.45\textwidth}{}}
\vspace{-2em}
\caption{Stokes~I image cutouts from the field of view near each star. The red cross-hairs highlight the star's location at the observing epoch (not detected).  The synthesized beam shape is shown in red in the bottom left.  The color scale is linear, with a low maximum set to highlight faint sources near the detection threshold. LTT~3780 \textit{(top left)} is a VLA 2-4~GHz image, whereas the remaining 5 are ATCA 1.1-3.1~GHz images.  GJ~367 has two separate images \textit{(center row)} for the epochs in January and September. The only Stokes~I source is LHS~3844 \textit{(top right)}, at $56\pm17$~\ujy in its second epoch (first epoch in main text).  LHS~1678 \textit{(bottom left)}, detected in Stokes~V (image in main text), has possible Stokes~I emission near the star's location, but it is offset from the star and not point-like, likely due to a background source or sidelobe.}
\label{fig:images}
\end{figure}

Figure~\ref{fig:images} shows Stokes~I clean images of the field of view around each star, zoomed in to the area near the star.
We cleaned a larger area of 1.5\degr\ to 3\degr\ (not shown).
The purpose of imaging was to create a model for the visibilities of background sources, subtracted in order to create the time series.  For most stars, we made a single image combining all epochs, since the epochs were separated by a few days, during which GHz-frequency background source variability is expected to be low.  The exceptions are GJ~367, for which we created separate images for January and September 2022, and LHS~3844, for which we imaged June 21 and 26 separately.

\section{Time Series} \label{app:tseries}

\begin{figure}[ht!]
\plotone{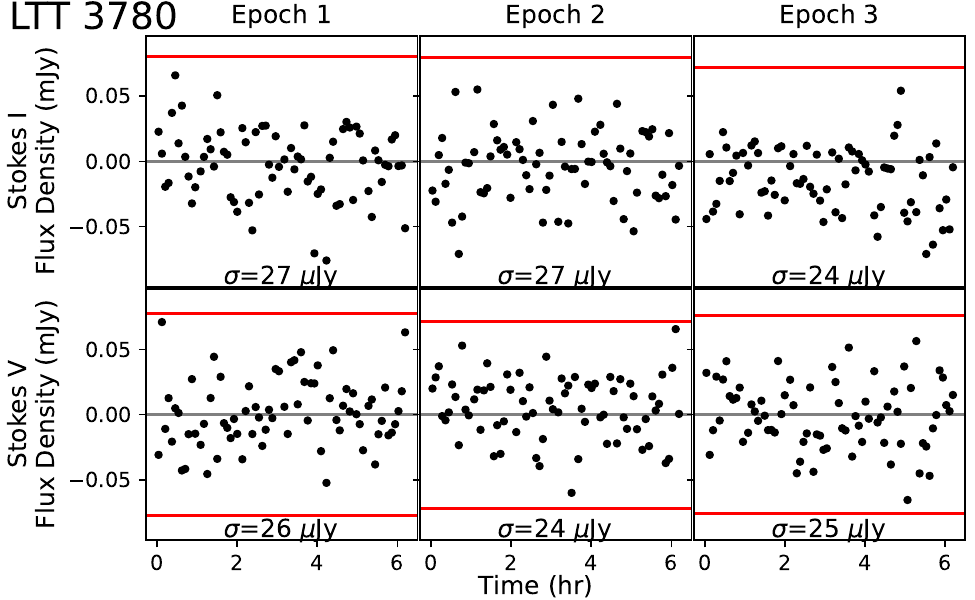}
\caption{VLA 2-4~GHz time series of LTT~3780. The black points show the flux density time series at the star's location, binned to 5-minute integrations, obtained by averaging the visibilities across all baselines. Red lines mark 3 times the standard deviation $\sigma$ of the time series.  No data exceeded the detection requirement of two or more adjacent points above $3\sigma$. Left to right: the epochs are the different observing dates in Table~\ref{tab:obs}. Top row: Stokes I flux density, where the detection threshold is positive only. Bottom: Stokes V, where the detection threshold can be positive (right-hand polarized emission) or negative (left-hand).
\label{fig:tseries1_LTT3780}}
\end{figure}

\begin{figure}[ht!]
\plotone{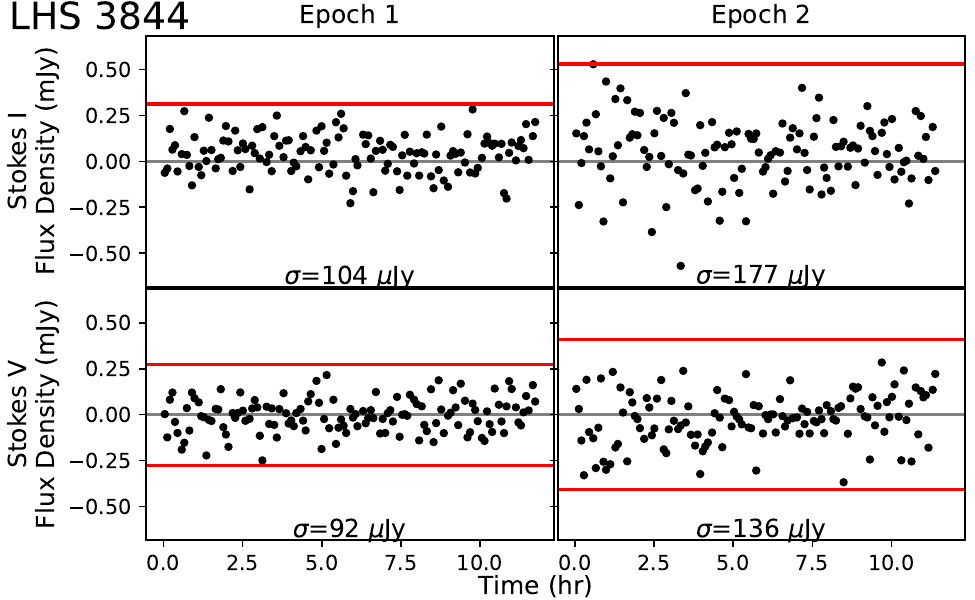}
\caption{ATCA 1.1-3.1~GHz time series of LHS~3844, in the style of Figure~\ref{fig:tseries1_LTT3780}.
\label{fig:tseries2_LHS3844}}
\end{figure}

\begin{figure}[ht!]
\plotone{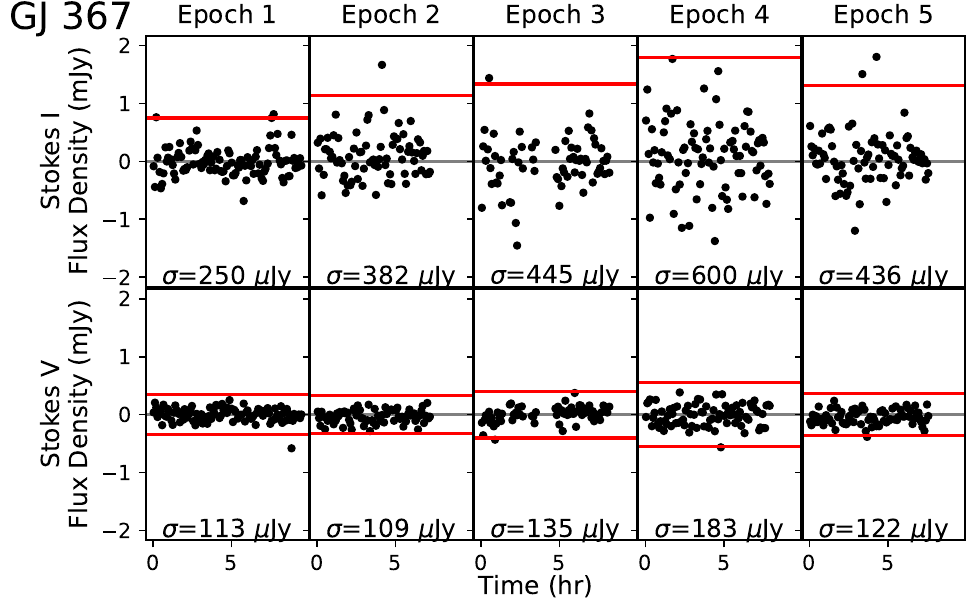}
\caption{ATCA 1.1-3.1~GHz time series of GJ~367, in the style of Figure~\ref{fig:tseries1_LTT3780}.
\label{fig:tseries3_GJ367}}
\end{figure}

\begin{figure}[ht!]
\plotone{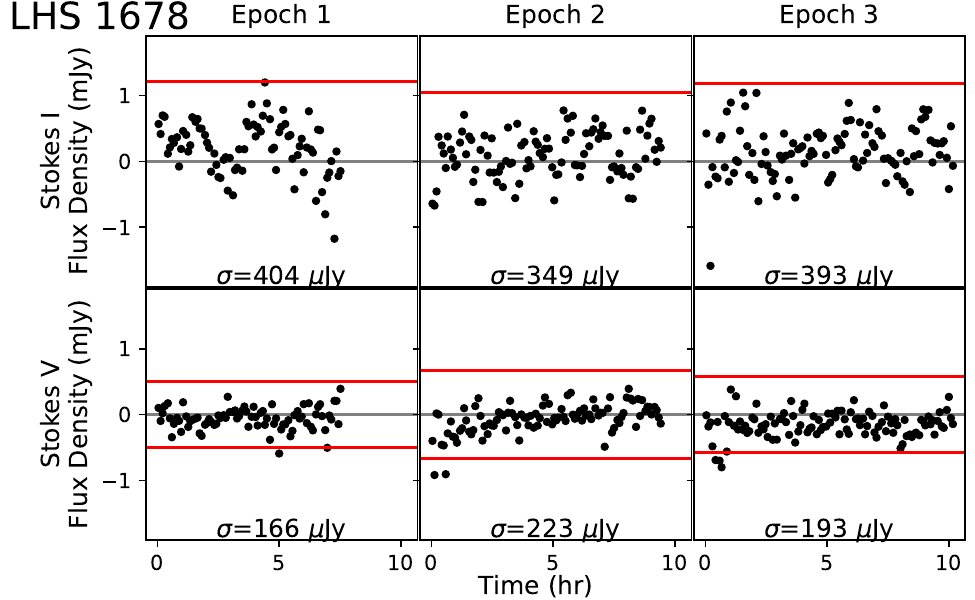}
\caption{ATCA 1.1-3.1~GHz time series of LHS~1678, in the style of Figure~\ref{fig:tseries1_LTT3780}.
\label{fig:tseries4_LHS1678}}
\end{figure}

\begin{figure}[ht!]
\plotone{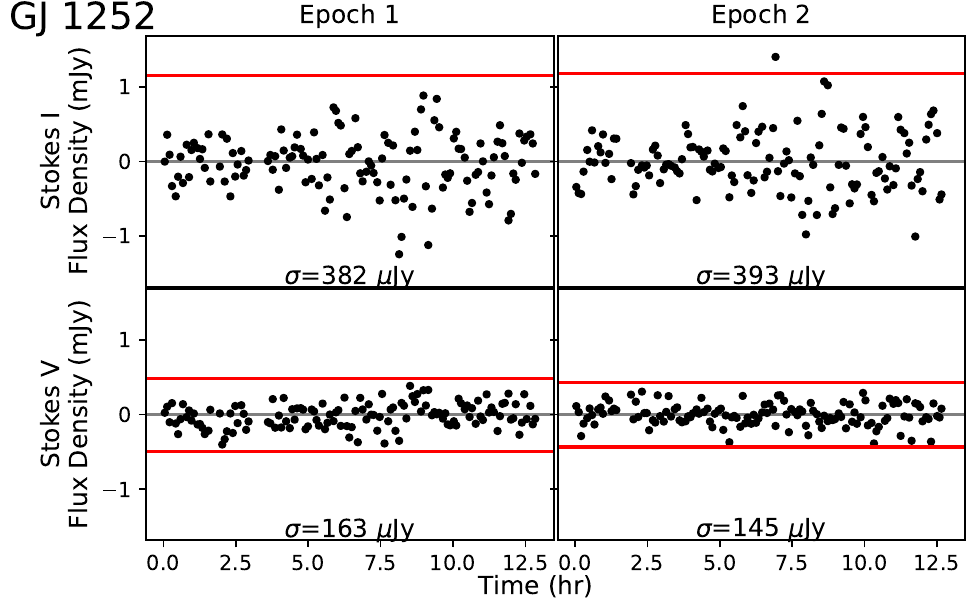}
\caption{ATCA 1.1-3.1~GHz time series of GJ~1252, in the style of Figure~\ref{fig:tseries1_LTT3780}.
\label{fig:tseries5_GJ1252}}
\end{figure}

Figures~\ref{fig:tseries1_LTT3780}-\ref{fig:tseries5_GJ1252} show the time series for the location of each star during the observing epochs, binned to 5-minute integrations. The $3\sigma$ thresholds were determined using the standard deviation $\sigma$ of each time series.  This works because the data reveal no clear stellar signal; the imaginary component of the visibilities (not shown), which should have no stellar signal, had comparable standard deviations.  Some individual points exceed a $3\sigma$ threshold, which can occur randomly due to the large number of data points plus non-Gaussian noise such as RFI, but no events meet the requirement of two or more adjacent points exceeding $3\sigma$.

\section{Simplifying $v \sin \theta$ for a Weber-Davis wind}
\label{app:v_sin_theta}

For a Weber-Davis wind model with an axisymmetric radial-azimuthal field, and a prograde circular orbit, then $v \sin \theta = \vorb (B_r/B) (1-\Porb/\Prot) \approx \vorb (1-\Porb/\Prot)$ for sub-Alfv\'enic systems, especially slow rotators. We demonstrate this approximation as follows.

The Weber-Davis wind model \citep{weberdavis1967wind} solves the combined flow problem for a magnetized equatorial wind around a rotating star. For star-planet interactions in this environment we needed to determine the Poynting flux in the exoplanet frame \citep[see][]{saur2013}. When assuming circular orbits of angular momentum aligned with the stellar rotational axis, we can make some useful approximations in this calculation. The key term is 

\begin{eqnarray}
| \vec{v} \times \vec{B} | = u_{r} B_{\phi} - (u_{\phi} - \vorb) B_r \, , \\
v B \sin \theta = u_{r} B_{\phi} - u_{\phi} B_{r} + B_{r} \vorb \, ,
\label{eq:vBcross}
\end{eqnarray}

\noindent where $v$ is the wind's relative velocity in the planet frame, $B$ is the wind magnetic field, $\theta$ is the angle between the two vectors in the planet frame, we use $u$ for the velocities of the wind, and $\vorb$ is the planet Keplerian orbital velocity. 

For the perfectly conducting wind, in the stellar rotating frame the plasma outflow is parallel to the magnetic field direction which further imposes the condition

\begin{equation}
    r (u_{r} B_{\phi} - u_{\phi} B_{r} ) = \mathrm{const.} = - \Omega r^{2} B_{r} \, ,
\end{equation}

\noindent where $\Omega$ is the angular rotation rate of the surface. The value of the constant is defined at the surface boundary at the roots of the wind, where the magnetic field is modeled as radial and the wind velocity as radial in the rotating frame. The constancy of the expression is a requirement of having a divergence less magnetic field ($\vec{\nabla} \cdot  \vec{B} = 0$).

This last expression lets us rewrite Equation~\ref{eq:vBcross} as

\noindent 

\begin{eqnarray}
    v B \sin \theta =  -\Omega r B_{r}+ B_{r} \vorb \, .
\end{eqnarray}

\noindent Rearranging, we can express 

\begin{equation}
v \sin \theta = \vorb \frac{ B_{r}}{B} \left( 1 - \frac{\Porb}{\Prot} \right)  \, ,
\end{equation}

\noindent which is an exact expression within the Weber-Davis model when applied to planets in circular orbits.

For a Weber-Davis wind, only the equatorial properties are included so the ratio of the radial field to the total field strength is 

\begin{equation}
\frac{B_{r}}{B} = \left( 1 + \frac{B_{\phi}^{2}}{B_{r}^{2}} \right)^{-1/2} \approx 1 - \frac{1}{2}\frac{B_{\phi}^{2}}{B_{r}^{2}} \, .
\end{equation}

\noindent In slowly rotating stars at close-in orbits $B_{\phi} << B_{r} $, so $B_{r} / B \approx 1$ to first order. In the Weber-Davis model the azimuthal wind follows

\begin{equation}
    \frac{B_{\phi}}{B_{r}} = - \frac{\Omega r}{u_{r,a}} \frac{ 1 - \frac{r^{2}}{r_{a}^{2}} }{ 1 - M_{A}^{2}} \, ,
\end{equation}

\noindent where $r$ is the radial position, the `$a$' subscripts are used to denote values evaluated at the Alfv\'{e}nic critical point (near the Alfv\'{e}n surface), and $M_{A}$ is the Alfv\'{e}nic Mach number, see \citet{weberdavis1967wind}. In a typical wind, the radial wind speed at the Alfv\'{e}nic critical point, $u_{r,a}$ , is several hundreds of kilometers per second. For the wind solution to exist, the fraction on the right is real and of order unity. Therefore, $B_{\phi} / B_{r}$ is small for the systems studied in this paper, and only becomes appreciable farther out when the stellar rotation period goes below 1~d: $\Omega r \sim$ 15 km s$^{-1}$ $(10~\mathrm{d} / \Prot) \, (r /  \, 10 R_{*}) $ for $R_{*} = 0.3 R_{\odot}$.

\end{document}